\documentclass[acmtog, nonacm]{acmart}

\AtBeginDocument{%
  \fancypagestyle{plain}{%
    \fancyhf{} %
    \fancyfoot[L]{ACM SIGENERGY Energy Informatics Review}%
    \fancyfoot[R]{Volume 1 Issue 1, November 2021}}
  \providecommand\BibTeX{{%
    \normalfont B\kern-0.5em{\scshape i\kern-0.25em b}\kern-0.8em\TeX}}}
    
\let\oldmaketitle\maketitle
\renewcommand{\maketitle}{%
  \oldmaketitle%
  \thispagestyle{plain}%
  \pagestyle{plain}}

\setcopyright{none}
\renewcommand\footnotetextcopyrightpermission[1]{}

\usepackage{tabularx,adjustbox}
\usepackage{graphicx}  
\usepackage{algorithm,algpseudocode} 
 \usepackage{longfbox}

\usepackage{amsmath}

\usepackage{array}
\usepackage{url}

\makeatletter
\newsavebox{\@brx}
\newcommand{\llangle}[1][]{\savebox{\@brx}{\(\m@th{#1\langle}\)}%
  \mathopen{\copy\@brx\kern-0.5\wd\@brx\usebox{\@brx}}}
\newcommand{\rrangle}[1][]{\savebox{\@brx}{\(\m@th{#1\rangle}\)}%
  \mathclose{\copy\@brx\kern-0.5\wd\@brx\usebox{\@brx}}}
\makeatother

\newcommand{\raf}[1]{(\ref{#1})}

\newtheorem{innercustomthm}{\textbf{Theorem}}
\newenvironment{customthm}[1]
{\renewcommand\theinnercustomthm{#1}\innercustomthm}
{\endinnercustomthm}

\usepackage{subcaption}
\begin{document}

\title{Privacy-Preserving Energy Storage Sharing with Blockchain and Secure Multi-Party Computation}

\author{Nan Wang}
\email{vincent.wang@anu.edu.au}
\affiliation{%
 \department{School of Computing}
  \institution{Australian National University}
}

\author{Sid Chi-Kin Chau}
\email{sid.chau@anu.edu.au}
\affiliation{%
 \department{School of Computing}
  \institution{Australian National University}
}

\author{Yue Zhou}
\email{yue.zhou@anu.edu.au}
\affiliation{%
 \department{School of Computing}
  \institution{Australian National University}
}

\renewcommand{\shortauthors}{N. Wang, S. C.-K. Chau, Y. Zhou}

\begin{abstract}
Energy storage provides an effective way of shifting temporal energy demands and supplies, which enables significant cost reduction under time-of-use energy pricing plans. Despite its promising benefits, the cost of present energy storage remains expensive, presenting a major obstacle to practical deployment. A more viable solution to improve the cost-effectiveness is by sharing energy storage, such as community sharing, cloud energy storage and peer-to-peer sharing. However, revealing private energy demand data to an external energy storage operator may compromise user privacy, and is susceptible to data misuses and breaches. In this paper, we explore a novel approach to support energy storage sharing with privacy protection, based on privacy-preserving blockchain and secure multi-party computation. We present an integrated solution to enable privacy-preserving energy storage sharing, such that energy storage service scheduling and cost-sharing can be attained without the knowledge of individual users' demands. It also supports auditing and verification by the grid operator via blockchain. Furthermore, our privacy-preserving solution can safeguard against a majority of dishonest users, who may collude in cheating, without requiring a trusted third-party. We implemented our solution as a smart contract on real-world Ethereum blockchain platform, and provided empirical evaluation in this paper\footnote{This article is an updated and extended version of the conference paper \cite{chau21blockchain}.}.

\end{abstract}


%\pagenumbering{gobble}

% remove ACM Reference after Abstract
%\settopmatter{printacmref=false}

\maketitle
\pagestyle{plain}

\section{Introduction}

\begin{figure*}[t]
\includegraphics[width=0.99\textwidth]{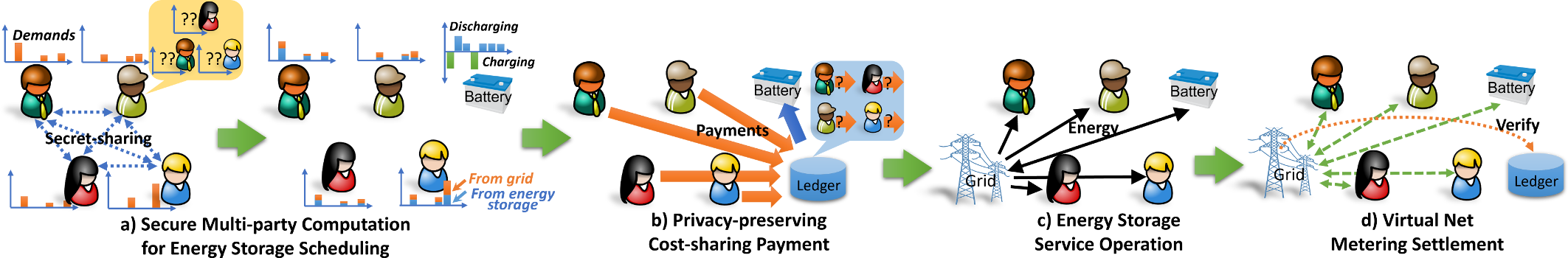}
\caption{An illustration of the stages of our solution for privacy-preserving energy storage sharing.}
\label{fig:scenario}
\end{figure*}

Energy storage can buffer energy in a storage medium, which is useful for temporal shifting of energy demands and supplies. In addition to absorbing excessive renewable energy, energy storage can effectively reduce the consumption cost under dynamic time-of-use (ToU) energy pricing plans by storing energy during off-peak periods and discharging during peak periods. But the present cost of energy storage systems remains considerably expensive. Energy storage also incurs significant maintenance cost over time, with only limited life cycles. 
There is a possibility of a future technological breakthrough that may significantly reduce the current cost of energy storage in the near future. Hence, despite its benefits, the current users are reluctant to immediately adopt energy storage at a wide scale. However, rather than postponing the use of energy storage, a more viable solution to improve the cost-effectiveness of present energy storage is by {\em sharing energy storage} among multiple users or out-sourcing to a third-party energy storage operator. In fact, time-sharing and out-sourcing have been popular concepts, particularly in cloud computing. Energy storage may also become an out-sourcible resource in a similar fashion. 

Currently, there are multiple possible paradigms of energy storage sharing. First, in {\em community sharing} \cite{vsolar18}, a group of local users, who do not own individual energy storage, can connect to a shared energy storage facility. The shared energy storage will be utilized by the users based on a coordination mechanism. The associated cost will be split among the users in a fair manner.  Second, a non-local third-party energy storage operator can provide an outsourcing service as {\em cloud energy storage} \cite{cloud17}. The energy storage operator can offset the energy consumption of remote users by exporting energy from its energy storage facility. Third, the users, who have their own energy storage, can pool their energy storage resources together to support each other in {\em peer-to-peer sharing} \cite{p2p19}.

All of these energy storage sharing paradigms can be effectively supported by the notion of {\em virtual net metering} (VNM) \cite{vnm20,vnmPGnE}, which is a flexible bill crediting system for transferring the credits or debits of a user's energy account to another, even though they do not share the same physical metering infrastructure. By VNM, energy storage operators can possibly transfer the credits of their energy export to offset the debits of energy import of other users. VNM has been used to enable community solar energy sharing in practice \cite{nrel14}. It can also enable energy storage sharing among a group of geographically distributed users and energy storage operators. 

Although sharing can improve the cost-effectiveness, there is a heightened concern of user privacy nowadays. Users may need to disclose private energy demand data to a third-party energy storage operator in order to schedule the use of shared energy storage. This may reveal sensitive personal data (e.g., working patterns, number of occupants, and vacation periods).  Potential misuses and breaches of personal data may lead to serious undesirable consequences. To bolster user privacy, stricter privacy protection legislations are being introduced in various countries to restrict personal information revelations to a third-party (e.g., GDPR in Europe). Because of these privacy concerns and privacy-related legislations, we are motivated to ensure proper privacy protection in energy storage sharing with a third-party operator. 

In this paper, we introduce the concept of ``{\em privacy-preserving energy storage sharing}'', by which a third-party energy storage operator should be given only minimal information for its energy storage service operations without being able to compromise personal data for other unintended purposes. But the key challenge is how to design an effective solution to enable proper energy storage service scheduling and cost-sharing among users, without the knowledge of individual users' energy demands, and yet that can still be verified and audited to eliminate any fraud.

We provide a feasible solution to enable privacy-preserving energy storage sharing, by drawing on several recent technologies. First, {\em blockchain} (e.g., Bitcoin, Ethereum) is a disruptive paradigm that enables decentralized verifiable applications without trusted intermediaries by integrating a tampering-resistant ledger with a distributed consensus protocol. Blockchain is an effective platform to support transparent energy storage sharing and auditable VNM with grid operators. But blockchain by default does not ensure privacy, and transaction data is entirely disclosed on the ledger. Recently, there is a new trend of supporting privacy on blockchain. For example, there are privacy-oriented cryptocurrencies, like ZCash, Monero \cite{zcash,monero}, that utilize zero-knowledge proofs for privacy-preserving digital asset management, without revealing them. In this paper, we utilize privacy-preserving blockchain to conceal the private data in cost-sharing and VNM for energy storage sharing. 

Second,  {\em secure multi-party computation} (or simply called multi-party computation) has been a subject of extensive research \cite{cramer2015secure}, which provides a general framework to allow multiple parties to jointly compute a function while concealing the private inputs. Recently, efficient multi-party computation protocols based on secret-sharing (e.g., SPDZ) have been applied to many practical applications like privacy-preserving machine learning \cite{cpr18spdz}. In this paper, we apply multi-party computation to energy storage service scheduling with concealed individual users' demands. Moreover, we integrate multi-party computation with privacy-preserving blockchain to support confidential cost-sharing and verifiable VNM settlement.

\smallskip

In summary, this paper presents an integrated solution to enable privacy-preserving energy storage sharing in all the stages, as outlined in the following (and also illustrated in Figure~\ref{fig:scenario}):
\begin{enumerate}

\item[(a)] {\bf Multi-party Computation for Energy Storage Scheduling}: First, the users can compute their aggregate day-ahead demands by secure multi-party computation, without revealing individual demands. Then, they can derive the optimal energy storage service schedule subject to energy storage service constraints. 

\item[(b)] {\bf Privacy-preserving Cost-sharing Payment}: The users can split the cost of energy storage service based on a fair cost-sharing scheme in a privacy-preserving manner. The users can make energy storage service payments via privacy-preserving blockchain, without disclosing individual transactions. After receiving the payments, the energy storage operator will issue verifiable receipts on blockchain ledger. 

\item[(c),(d)] {\bf Energy Storage Operation \& Virtual Net Metering Settlement}: The users and energy storage operator will follow the energy storage service schedule. They do not need to exchange energy directly, and the energy flows through the grid. They will settle their energy accounts via VNM. With verifiable receipts on blockchain ledger, the users can offset their energy consumption by the energy export from energy storage, which will be audited by the grid operator. 

\end{enumerate}
Particularly, we should ensure privacy protection throughout the {\em integrated} process of scheduling, cost-sharing, payment and auditing, without requiring a trusted third-party. While it may be easier to ensure privacy in individual processes separately, it is challenging to ensure privacy in the integrated process. For example, one can schedule a service, or make a payment separately in a privacy-preserving manner. However, it is harder to verify the payment with respect to the scheduled service with privacy protection.

Furthermore, privacy also poses a significant challenge to the correctness and integrity of operations. Because of concealing their demands, dishonest users may attempt to cheat by paying less to energy storage service or claim more in VNM than what they ought to. These dishonest users may even collude to coordinate their actions in cheating. Hence, it is critical to safeguard against the presence of dishonest users. Remarkably, our privacy-preserving solution is able to safeguard against a {\em majority of dishonest users} (namely, more than 50\% of users may be dishonest).

\medskip

This paper is organized as follows. We first review the related work and background in Section~\ref{sec:related}. We then formulate the problem and models in Section~\ref{sec:model}, and present the basics of cryptographic components and multi-party computation in Sections~\ref{sec:crypto}-\ref{sec:spdz}. The privacy-preserving solution is presented in Section~\ref{sec:algo}. We next provide an empirical evaluation of our implementation on Ethereum blockchain platform in Section~\ref{sec:eval}. We also discuss several extensions as well as the limitations of our solution in Section~\ref{sec:extension}. We conclude this work with future work in Section~\ref{sec:concl}.

\section{Related Work and Background} \label{sec:related}

\subsection{Energy Storage Sharing}

Optimizing energy storage under dynamic pricing plans has been a popular research topic \cite{crowd17,control14, smartCharge12}. Recent studies proposed various paradigms for energy storage sharing among multiple users, for instance, cloud energy storage \cite{cloud17}, virtual community sharing \cite{vsolar18} and peer-to-peer sharing \cite{p2p19}. Notably, there are many studies about privacy in smart grid in other aspects. For example, \cite{privbat16, shepherd16} employed energy storage to hide private consumption behavior by mixing random energy storage charging and discharging to mask the consumption patterns. \cite{shi2011privacy} presents privacy-preserving data aggregation for smart meters that aggregates users demands. None of these studies addressed the privacy aspect in energy storage sharing. To the best of our knowledge, this is the first paper to address the issue of privacy-preserving energy storage sharing and its cost-sharing.

\subsection{Virtual Net Metering}

To enable energy consumers to share physically disconnected energy storage from energy storage operators, one can rely on Virtual Net Metering (VNM) \cite{vnm20,vnmPGnE} for transferring the credits or debits of a user's energy account to another. When energy consumers import energy from the grid, they will incur debits in their energy accounts. On the other hand, when energy storage operators export energy to the grid, they will earn credits in their energy accounts based on feed-in tariffs. However, simultaneous exporting energy from energy storage operators and importing energy from energy consumers with the same amount of energy should be able to offset each other. 

In practice, the credits of energy export of energy storage operators may be transferred to offset the partial debits of energy import of energy consumers. In this case, it may not require simultaneous energy export and energy import in VNM. However, maintaining instantaneous energy balance at VNM is still important to ensure the stability in power distribution network. While there may be additional costs in power distribution network such as power transmission cost, balancing the energy generation and consumption should be the major component in VNM. 
Note that VNM is entirely an account balancing process, without the need to configure the energy flow in the power distribution network. VNM has been proposed to enable novel applications, such as transactive energy transfer in an energy exchange market, community solar energy and shared energy storage \cite{nrel14}.

\subsection{Blockchain Technology}

There is an increasing number of applications of blockchain technology to energy systems. For example, the study \cite{ggk19block} applied blockchain to mitigate trust in peer-to-peer electric vehicle charging. Blockchain has been applied to microgrid energy exchange and wholesale markets by prosumers \cite{m18brooklyn}. Renewable energy credits and emissions trading are also applications of blockchain \cite{kbue20gecko}. In these applications, the goal of blockchain is to improve transparency and reduce settlement times, since blockchain system can ensure integrity and consistency of transactions and settlement on an open ledger. See \cite{a19survey} for a recent survey about blockchain applications to energy systems. 

Note that none of these studies have considered the privacy on blockchain, even though the transaction data on the ledger is entirely disclosed to the public. Our work is one of the first studies to explicitly address privacy in blockchain applications of energy systems. Supporting privacy on blockchain is a crucial research topic in cryptography and security. There have been several privacy-preserving blockchain platforms with support of privacy (e.g. ZCash, Monero, Zether \cite{zcash,monero,bunz20zether}). Our work draws on similar concepts from privacy-preserving blockchain, but also integrates specifically with the application of energy storage sharing, for example, to support auditable VNM. Our solution is implemented as a smart contract on permissionless Ethereum blockchain platform, but it can also be implemented on a permissioned blockchain platform.

\subsection{Privacy-Preserving Solutions}

We briefly survey and compare various approaches of privacy-preserving solutions in the literature. There are two major approaches: (1) data obfuscation that masks private data with random noise, (2) secure multi-party computation that hides private data while allowing the data to be computed confidentially. Differential privacy \cite{dwork2006calibrating}, a main example of data obfuscation, is often used in privacy-preserving data mining to extract certain data properties in a relatively large dataset. There is an intrinsic trade-off between the accuracy and privacy of differential privacy. On the other hand, secure multi-party computation \cite{goldreich1998secure, du2001secure} traditionally employed garbled circuits \cite{HazayL10} and homomorphic cryptosystems \cite{cramer2001multiparty,lcwz20privsharing}, which have a high computational complexity. Recently, information-theoretical secret-sharing (e.g., SPDZ \cite{cramer2015secure, dklpss13spdz}) has been utilized for secure multi-party computation, which provides high efficiency. This work employs secure multi-party computation for privacy-preserving energy storage operation scheduling and cost-sharing computations without disclosing private energy demands.

\section{Models and Formulation}  \label{sec:model}

In the following, we first formulate the energy storage sharing model without considering privacy. In the subsequent sections, we will incorporate privacy protection in the model.

\subsection{Problem Setup} 

First, we describe several key components in the model (and list some key notations in Table~\ref{tbl:symbs}):
%%%%%%%%%%%%%%%%%%%%%%%%%%%%%%%%%%%%%%%
\begin{table}[!h]
    \centering
    \caption{\label{tbl:symbs}Table of key symbols and notations.} 
    \begin{tabularx}{\linewidth}{@{}c@{ }X@{}}
    \hline  \hline   
   $N$  & Total number of users \\
   $U_i$  & The $i$-th user  \\
   %$T$  & Total number of timeslots \\
   $p(t)$  & Energy price of time-varying pricing scheme at timeslot $t$ \\
   ${\tt B}(t)$ & Capacity of energy storage at timeslot $t$\\   
   ${\tt p}_{\tt s}$ & Per-unit service fee of energy storage at each timeslot\\  
   ${b}(t)$  & State-of-charge of energy storage at timeslot $t$ \\
   ${\tt e}_{\tt c}, {\tt e}_{\tt d} $ & Charging and discharging efficiency ratios \\
   ${\tt r}_{\tt c}, {\tt r}_{\tt d} $ & Charging and discharging rate constraints \\
   ${x}^+(t)$  & Charging rate from the grid to the energy storage \\
   ${x}^-_i(t)$  & Discharging rate from the energy storage to $U_i$ \\
   ${y}_i(t)$  & $U_i$'s residual consumption rate from the grid at $t$\\
   ${y}(t)$  &  Total residual consumption rate of all users at timeslot $t$ \\   
   ${\tt Cost}_{\tt ess}$ & Total cost of energy storage service \\
   ${\tt Cost}_i$ & $U_i$'s partial original cost without energy storage service \\
   $P_i^{\tt pp}$ & $U_i$'s payment under proportional cost-sharing scheme \\
   $P_i^{\tt ega}$ & $U_i$'s payment under egalitarian cost-sharing scheme \\
   $\Delta_i$ & $(={\tt Cost}_i - P_i)$ $U_i$'s saving from energy storage service\\
    \hline  \hline    
    \end{tabularx}
\end{table}
%%%%%%%%%%%%%%%%%%%%%%%%%%%%%%%%%%%%%%%

\begin{enumerate}

\item {\bf Time-Varying Energy Pricing Plan:} 
We consider discrete timeslots, indexed by $t \in \{1, ..., T \}$, where $T$ is the number of timeslots in a day. The energy price of a time-varying time-of-use (ToU) pricing plan at timeslot $t$ is denoted by $p(t)$. We suppose that the next-day ToU prices $(p(t))_{t=1}^T$ are announced before the end of today to all users and energy storage operator, such that they can plan their consumption in a day-ahead manner.

\item {\bf Energy Users:}  There are $N$ ($\ge 3$)\footnote{Our protocols can safeguard against at most $N-2$ dishonest users.} users, each denoted by $U_i$ where $i \in \{1, ..., N\}$. $U_i$ has certain energy demand over time, represented by a non-negative demand function $a_i(t) > 0$ for all $t$. The users aim to reduce their energy costs by utilizing a third-party energy storage service that stores energy at lower energy prices beforehand. We consider day-ahead energy storage scheduling, whereby $U_i$ forecasts her planned energy demand $a_i(t)$ in advance, and requests energy storage service in a day-ahead manner. If the energy from energy storage service is insufficient, $U_i$ will need to acquire additional energy from the grid for the residual consumption rate denoted by ${y}_i(t)$ at the respective price $p(t)$.

\item {\bf Energy Storage Service:} The energy storage service is provided by an energy storage operator, who has energy storage characterized by capacity ${\tt B}(t)$, which is time-varying for modeling dynamic energy storage capacity. The energy storage is constrained by charging efficiency ratio ${\tt e}_{\tt c} \le 1$ and discharging efficiency ratio ${\tt e}_{\tt d} \ge 1$, charge rate (i.e., ramp-up) constraint ${\tt r}_{\tt c}$ and discharge rate (i.e., ramp-down) constraint ${\tt r}_{\tt d}$. Let $b(t) $ be the current state-of-charge in the energy storage at time $t$, and ${x}^+(t)$ be the charging rate from the grid to the energy storage, whereas ${x}^-_i(t)$ be the discharging rate from the energy storage to $U_i$. When the energy storage is utilized, there is a per-unit service fee at each timeslot, ${\tt p}_{\tt s}$, which allows the energy storage operator to cover the wear-and-tear and maintenance cost.

\end{enumerate}

Next, we will describe energy storage service scheduling in Section~\ref{sec:model-schedule} and fair energy storage service cost-sharing in Section~\ref{sec:model-costsharing}. We will present the blockchain model in Section~\ref{sec:model-blockchain}, and incorporate privacy protection in the security and threat models in Section~\ref{sec:model-security}.

\subsection{Energy Storage Service Scheduling}  \label{sec:model-schedule}

The energy storage service requires reservations from the users. The energy storage service operations will then be scheduled accordingly to minimize the overall energy cost. We formulate the optimization problem of energy storage service scheduling in ${\tt (P1)}$.
\begin{align} 
{\tt (P1)} \ & \ \min  \sum_{t=1}^{T} \Big( p(t) \cdot \big({x}^+(t)+ \sum_{i=1}^N {y}_i(t) \big) +  {\tt p}_{\tt s} \cdot {x}^+(t) \Big) \label{eqn:objective}\\
\textrm{s.t.\ } &   b(t+1) - b(t)  =  {\tt e}_{\tt c} {x}^+(t)- {\tt e}_{\tt d} \Big(\sum_{i=1}^N {x}^-_i(t) \Big), & \label{eqn:SoC}\\
 &  0 \le b(t) \le {\tt B}(t), b(0)=0, b(T+1)=0,  &  \label{eqn:over-underflow}\\
 &   {x}^+(t)  \le {\tt r}_{\tt c},  &  \label{eqn:charge-rate}\\ 
 &   \sum_{i=1}^N {x}^-_i(t)  \le {\tt r}_{\tt d},  &  \label{eqn:discharge-rate}\\ 
 &  {x}^-_i(t) +  {y}_i(t)   = a_i(t),  &  \label{eqn:i-demand-balance}\\
 \textrm{var. \ } & b(t) \ge 0, {x}^-_i(t)\ge 0,  {y}_i(t)\ge 0, {x}^+(t) \ge 0  \\
 & \forall t \in \{1,...,T\}, \forall i \in \{1,...,N\}    \notag 
\end{align}

The objective of ${\tt (P1)}$ is the total cost, including energy storage charging ${x}^+(t)$ and residual consumption $ {y}_i(t)$ at the respective energy price $p(t)$ of timeslot $t$, as well as the energy storage service fee ${\tt p}_{\tt s} \cdot {x}^+(t)$. Constraint~\raf{eqn:SoC} updates the state-of-charge considering charging and discharging efficiency ratios. Constraint~\raf{eqn:over-underflow} ensures feasible state-of-charge. We assume that the initial and final state-of-charge are 0. Constraints~\raf{eqn:charge-rate}-\raf{eqn:discharge-rate} ensure the charging and discharging rates within the respective rate constraints. Constraint~\raf{eqn:i-demand-balance} ensures the balance of demands, such that each user's demands are satisfied completely. Note that we do not consider the cost of energy distribution in power distribution network. This is sufficient to certain scenarios, for example, when the users are close to the energy storage operator.

We note that ${\tt (P1)}$ however relies on the knowledge of individual user's demand $a_i(t)$. Hence, we present an alternate problem ${\tt (P2)}$.
\begin{align} 
{\tt (P2)} \ & \ \min  \sum_{t=1}^{T} \Big( p(t) \cdot \big({x}^+(t)+  {y}(t) \big) +  {\tt p}_{\tt s} \cdot {x}^+(t) \Big) \\
\textrm{s.t.\ } &   b(t+1) - b(t)  =  {\tt e}_{\tt c} {x}^+(t)- {\tt e}_{\tt d}  {x}^-(t), & \\
 &  0 \le b(t) \le {\tt B}(t), b(0)=0, b(T+1)=0, &  \\
 &   {x}^+(t)  \le {\tt r}_{\tt c},  &  \\ 
 &   {x}^-(t)  \le {\tt r}_{\tt d},  &  \\ 
 &  {x}^-(t) +  {y}(t)   = a(t),  & \label{eqn:demand-balance}\\
 \textrm{var. \ } & b(t) \ge 0, {x}^-(t)\ge 0,  {y}(t)\ge 0, {x}^+(t) \ge 0  & \forall t \in \{1,...,T\} \notag 
\end{align}
${\tt (P2)}$ considers the total demand  ${a}(t)  =\sum_{i=1}^N a_i(t)$, total discharging rate ${x}^-(t) \triangleq \sum_{i=1}^N {x}^-_i(t)$ and total consumption rate ${y}(t)  = \sum_{i=1}^N {y}_i(t)$, as well as the balance of the total demand in Constraint~\raf{eqn:demand-balance}.
By Theorem~1, energy storage service scheduling can be solved by ${\tt (P2)}$, instead of ${\tt (P1)}$, involving no individual demand ${a}_i(t)$. 

\begin{customthm}{1} 
If $\big({x}^-(t), {y}(t)\big)_{t=1}^T$ is an optimal solution of ${\tt (P2)}$, then $\big(({x}^-_i(t), {y}_i(t))_{i=1}^N\big)_{t=1}^T$, where ${x}^-_i(t) =  \frac{a_i(t)}{a(t)} \cdot {x}^-(t)$ and ${y}_i(t) = \frac{a_i(t)}{a(t)} \cdot {y}(t)$, is an optimal solution of ${\tt (P1)}$.
\end{customthm}

See Appendix.~\ref{sec:append1} for the proof. 

\smallskip

{\bf Remarks}: Note that when the energy storage discharges at rate ${x}^-(t)$, it can simultaneously compensate the users' consumption at the same rate. This can be attained via VNM. 
%Also, the users are not supposed to alter their planned demands in the subsequent operation stage. But, we will discuss extensions accommodate flexible scheduling later in Section~\ref{sec:extension}. 
We assume that the energy storage operator announces the parameters ${\tt p}_{\tt s}, {\tt e}_{\tt c},$ ${\tt e}_{\tt d}, {\tt r}_{\tt c}, {\tt r}_{\tt d}, (p(t),$\\ ${\tt B}(t))_{t=1}^T$ in advance. Everyone can compute the solution to ${\tt (P2)}$ with the knowledge of  $(a(t))_{t=1}^T$.

\subsection{Fair Cost-sharing of Energy Storage Service}  \label{sec:model-costsharing}

After scheduling the energy storage service, the users are supposed to share and pay the associate cost to the energy storage operator. Next, we formulate how the cost of energy storage service should be shared among users in a fair manner. 
 In ${\tt (P2)}$, in addition to the cost that is paid directly by the users to the grid (i.e., $\sum_{t=1}^T p(t) \cdot {y}(t)$), there is a cost incurred by the energy storage service as follows:
\begin{equation}
{\tt Cost}_{\tt ess} \triangleq \sum_{t=1}^T \big( p(t) + {\tt p}_{\tt s} \big) \cdot {x}^+(t)
\end{equation}
${\tt Cost}_{\tt ess}$ should be split fairly among the users. There are several possible ways of dividing  the energy storage service cost ${\tt Cost}_{\tt ess}$. Particularly, we are interested in the fair ways that take into consideration the individual rationality of each user.

We note that without energy storage service, each user should originally pay the following cost:
\begin{equation}
\sum_{t=1}^T p(t) \cdot {a}_i(t) = \sum_{t=1}^T p(t) \cdot \big( {x}^-_i(t) +  {y}_i(t) \big) \label{eqn:orgcost}
\end{equation}

Let ${\tt Cost}_i \triangleq \sum_{t=1}^T p(t) \cdot {x}^-_i(t)$ be the partial original cost of $U_i$ in Eqn.~\raf{eqn:orgcost} that would have been covered by energy storage service, which provides a basis on how to split ${\tt Cost}_{\tt ess}$. Note that the other part in Eqn.~\raf{eqn:orgcost} (i.e., $\sum_{t=1}^T p(t) \cdot {y}_i(t)$) will be paid regardless of energy storage service. Noteworthily, if a user does not get any benefit from energy storage service (i.e., ${a}_i(t) > 0$ only when $p(t)$ is the lowest), then we have ${x}^-_i(t) =0$ in $({\tt P1})$ and ${\tt Cost}_i = 0$.

Suppose that each $U_i$ contributes payment $P_i$ to cover the energy storage service cost ${\tt Cost}_{\tt ess}$. A cost-sharing scheme denoted by $(P_i)_{i=1}^N$ is called {\em budget-balanced}, if $\sum_{i=1}^N P_i = {\tt Cost}_{\tt ess}$, whereas it is called {\em weakly budget-balanced}, if $\sum_{i=1}^N P_i \ge {\tt Cost}_{\tt ess}$. A cost-sharing scheme $(P_i)_{i=1}^N$ is called {\em individually rational}, if ${\tt Cost}_i \ge P_i$ for all $i \in \{1, ..., N\}$. Evidently, each user would prefer an individually rational cost-sharing scheme. Otherwise, some users would rather not to utilize energy storage service, as it will cost more.

We define two  fair cost-sharing schemes, which are based on similar ideas in \cite{p2p19,CE17sharing}, and show them to be individually rational by Theorem~2.

\subsubsection{Proportional Cost-sharing Scheme}
\

One simple fair way is that each $U_i$ should pay proportionally to ${\tt Cost}_i$. Namely,
$$
P_i^{\tt pp} \triangleq {\tt Cost}_{\tt ess} \cdot \frac{{\tt Cost}_i}{\sum_{i=1}^N {\tt Cost}_i} = \sum_{t=1}^T \big( p(t) + {\tt p}_{\tt s} \big) \cdot {x}^+(t) \cdot \frac{\sum_{t=1}^T {x}^-_i(t) \cdot p(t)}{\sum_{t=1}^T {x}^-(t) \cdot p(t)}
$$
Thus, each user has the same ratio of payment over individual cost (i.e.,  $\frac{P_i^{\tt pp} }{{\tt Cost}_i} = \frac{{\tt Cost}_{\tt ess}}{\sum_{i=1}^N {\tt Cost}_i}$). It is easy to check that proportional cost-sharing is budget-balanced (i.e., $\sum_{i=1}^N P_i^{\tt pp} = {\tt Cost}_{\tt ess}$). Note that the payments are always non-negative (i.e., $P_i^{\tt pp} \ge 0$).

\subsubsection{Egalitarian Cost-sharing Scheme}
\

Given a payment to energy storage service $P_i$, define the user's {\em saving} of utilizing energy storage service by $\Delta_i \triangleq {\tt Cost}_i - P_i$. Another fair cost-sharing scheme is that each user should split ${\tt Cost}_{\tt ess}$  in a way that attains the same saving for every user. Namely,
\begin{align}
 P_i^{\tt ega} & \triangleq  {\tt Cost}_i - \frac{\sum_{i=1}^N {\tt Cost}_i - {\tt Cost}_{\tt ess}}{N} \\
& = \sum_{t=1}^T {x}^-_i(t) \cdot p(t) - \frac{\sum_{t=1}^T {x}^-(t) \cdot p(t) - \sum_{t=1}^T \big( p(t) + {\tt p}_{\tt s} \big) \cdot {x}^+(t)}{N} \notag
\end{align}
Thus, each $U_i$ attains the same saving as: $\Delta_i^{\tt ega} \triangleq \frac{\sum_{i=1}^N {\tt Cost}_i - {\tt Cost}_{\tt ess}}{N}$.
It is easy to check that egalitarian cost-sharing is also budget-balanced (i.e., $\sum_{i=1}^N P_i^{\tt ega} = {\tt Cost}_{\tt ess}$).

\smallskip
As a comparison, proportional cost-sharing guarantees the same percentage of savings (i.e., $\frac{\Delta_i}{{\tt Cost}_i }$) among users, whereas egalitarian cost-sharing guarantees the same savings (i.e., ${\Delta_i}$) among users. 

\begin{customthm}{2} \label{lem:rational}
If $\big({x}^+(t), {x}^-(t)\big)_{t=1}^T$ is an optimal solution of ${\tt (P2)}$ and let ${x}^-_i(t) =  \frac{a_i(t)}{a(t)} \cdot {x}^-(t)$ and ${y}_i(t) = \frac{a_i(t)}{a(t)} \cdot {y}(t)$, then proportional and egalitarian cost-sharing schemes are individually rational.

Let $\hat{\tt p}(t) \triangleq \frac{{x}^-(t) \cdot p(t)}{a(t)}$ and ${\tt Cost}_{\tt org} \triangleq \sum_{t=1}^T {x}^-(t) \cdot p(t)$. The proportional and egalitarian cost-sharing payments are given as follows:
\begin{equation}
\begin{cases}
P_i^{\tt pp} =  \frac{{\tt Cost}_{\tt ess}}{{\tt Cost}_{\tt org}} \cdot \sum_{t=1}^T a_i(t) \cdot \hat{\tt p}(t), \\  
P_i^{\tt ega} =  \sum_{t=1}^T a_i(t) \cdot \hat{\tt p}(t)  - \frac{{\tt Cost}_{\tt org}  - {\tt Cost}_{\tt ess}}{N} 
\end{cases}
\end{equation}
\end{customthm}

See Appendix.~\ref{sec:append1} for the proof. 

\smallskip

{\bf Remarks}: Egalitarian cost-sharing may have negative payments (i.e., $P_i^{\tt ega} < 0$), when ${\tt Cost}_i < \Delta_i^{\tt ega}$. Namely, a user may be paid by other users who have larger original costs, in order to maintain equal savings among all users. In this case, such a user is not benefited sufficiently from energy storage service because of the presence of other users and capacity constraint, and hence, will be compensated by other users in egalitarian cost-sharing. 

One may argue whether proportional cost-sharing is better than egalitarian cost-sharing, because it rules out negative payments. Here, we provide a solution to support both cost-sharing schemes. We will leave the decision of adopting which scheme to the users. 
	
\subsection{Blockchain Model} \label{sec:model-blockchain}

In this section, we describe a blockchain model for payments of energy storage service. We consider an account-based blockchain model like Ethereum (which is a general-purpose blockchain platform \cite{ethereum}), whereas Bitcoin operates with a different transaction-output-based model for cryptocurrency transactions only. Smart contracts are programming code on a blockchain that can provide customized computation tasks to each transaction (e.g., verification, data processing). Our payment system can be implemented as a smart contract. 

The payment and auditing of energy storage service are carried out on a blockchain. Each user has an account on the blockchain. Users can top-up their accounts in advance. For cost-sharing, the users can initiate a joint payment transaction to the energy storage operator. The transaction records on the blockchain will also be used to verify VNM settlement by the grid operator.

Our blockchain model is based on a common model in the cryptography literature (e.g., Zether \cite{bunz20zether} that was built on Ethereum), which can be incorporated with privacy protection to conceal the transaction records. The blockchain consists of several components:

\begin{enumerate}

\item {\bf Ledger}: An append-only ledger on a blockchain holds the records of all accounts and transactions. Note that by default, there is no privacy protection to the ledger, such that the account details and transaction histories are visible to the public. On Ethereum, one can create tokens on the ledger to represent certain digital assets. Our payment system is implemented by tokens, which allows us to incorporate privacy protection. To pay for energy storage service, users are required to purchase tokens that will be subsequently transfered to the energy storage operator and redeemed.  

\item {\bf Accounts}: An account is identified by a public key $K^{\tt p}$ and an address ${\tt ad}$, which is the hash of the public key: ${\tt ad} = {\mathcal H}(K^{\tt p})$, where ${\mathcal H}(\cdot)$ is a cryptographic hash function. The user manages the account by a private key $K^{\tt s}$. Each account holds a balance of tokens, denoted by ${\tt Bal}({\tt ad})$, which by default is a publicly visible plaintext. Each $U_i$ has an account associated with a tuple $({\tt ad}_i, K^{\tt p}_i, K^{\tt s}_i, {\tt Bal}({\tt ad}_i))$. We denote the energy storage operator's account address by ${\tt ad}_{\tt ess}$.

\item {\bf Transactions}: To initiate a transaction of tokens from ${\tt ad}_i$ to ${\tt ad}_{i'}$ with transaction value ${\tt val}$, the user submits a transaction request to the blockchain: ${\tt tx} = ({\tt ad}_i, {\tt ad}_{i'}, {\tt val})$, along with a signature ${\tt sign}_{K^{\tt s}_i}({\tt tx})$ using the private key $K^{\tt s}_i$ associated with ${\tt ad}_i$. The transaction request will be executed\footnote{We skip some practical issues of a blockchain transaction, like nonce to prevent replay attack, account-locking against front-running attack, etc. But our model can easily incorporate the solutions from the security literature (e.g., \cite{bunz20zether}) to address these issues.} if ${\tt Bal}({{\tt ad}_i}) \ge {\tt val}$. A multi-transaction can also be requested. Let ${\tt mtx} = ({\tt ad}_i, {\tt ad}_{i'}, {\tt val}_i)_{i=1}^N$. ${\tt mtx}$ will be executed, only if ${\tt Bal}({\tt ad}_i) \ge {\tt val}_i$ for all $i$ and multi-signature ${\tt sign}_{(K^{\tt s}_i)_{i=1}^N}({\tt mtx})$ is present. Depending on the cost-sharing scheme, a user will pay either $P_i^{\tt pp}$ or $P_i^{\tt ega}$ to the energy storage operator. Each transaction request by default is a plaintext visible to the public. We will subsequently conceal the transaction records.

\item {\bf Receipts}: The recipient of a transaction can attach a receipt on the ledger, which may include additional information for further verification and auditing by a third-party. In VNM settlement, the grid operator will need to audit the amount of energy that a user can be offset from energy storage service, which can be verified from the receipts associated with transaction records.

\end{enumerate}

Note that there may be a negative flow of payment in egalitarian cost-sharing, such that ${\tt val}_i < 0$. Hence, we need to ensure the corresponding transaction on a blockchain still functions correctly. 

\begin{customthm}{3} \label{lem:negflow}
Consider a multi-transaction ${\tt mtx} = ({\tt ad}_i, {\tt ad}_{\tt ess},$ ${\tt val}_i)_{i=1}^N$, where ${\tt val}_i$ may be negative. Namely, every ${\tt ad}_i$ pays to the energy storage operator ${\tt ad}_{\tt ess}$. If $\sum_{i=1}^N {\tt val}_i > 0$, then ${\tt mtx}$ can be handled on a blockchain by the following transaction operations:
\begin{align}
{\tt Bal}({\tt ad}_{i}) \leftarrow & {\tt Bal}({\tt ad}_{i}) - {\tt val}_i, \mbox{for all\ }i \\
{\tt Bal}({\tt ad}_{\tt ess}) \leftarrow & {\tt Bal}({\tt ad}_{\tt ess}) + \sum_{i=1}^N {\tt val}_i
\end{align}
\end{customthm}

See Appendix.~\ref{sec:append1} for the proof.

\subsection{Security \& Threat Models}  \label{sec:model-security}

In the previous sections, we have not considered privacy protection. We define privacy protection in our problem. We assume synchronously authenticated communications among the parties, including users, blockchain, energy storage operator and grid operator, where the protocols proceed in several rounds and the parties can authenticate each other properly so that there is no man-in-the-middle attack.

\subsubsection{Security Requirements} \label{sec:req}
\

Our system aims to satisfy the following security requirements:
\begin{enumerate}

\item[({\tt S1})] {\bf Demand Concealment}:
The user's demand $\big(a_i(t)\big)_{t=1}^T$ is private information, which should not be revealed to other users or energy storage operator in energy storage service scheduling, cost-sharing and payment. But the parameters, such as ${\tt p}_{\tt s}, {\tt e}_{\tt c}, {\tt e}_{\tt d},$ ${\tt r}_{\tt c}, {\tt r}_{\tt d}, \big(p(t), {\tt B}(t)\big)_{t=1}^T$,  are publicly known to all users. We need to ensure the operations of scheduling, cost-sharing and payment can be achieved correctly without leaking any information about $\big(a_i(t)\big)_{t=1}^T$ to others. Specifically, given $\Big(\big(a_i(t)\big)_{t=1}^T\Big)_{i=1}^N$, we need a privacy-preserving summation function for the aggregate demand:
\[
{\tt Sum}_{\tt prv} \Big[ \Big(\big(a_i(t)\big)_{t=1}^T\Big)_{i=1}^N \Big] = \big(a(t)\big)_{t=1}^T.
\]
No user should learn any information from ${\tt Sum}_{\tt prv}[\cdot]$ other than her own inputs and the final outputs.

\item[({\tt S2})] {\bf Zero-knowledge Cost-Sharing \& Payment}:
With $\big(a(t)\big)_{t=1}^T$, one can compute the energy storage service schedule $\big(({x}^+(t),$ ${x}^-(t), {y}(t)\big)_{t=1}^T$ by Theorem~1. Then, each $U_i$ can compute and make her payment $P_i (= P_i^{\tt pp} \mbox{\ or\ } P_i^{\tt ega})$ by Theorem~2.
Since $a_i(t)$ is only known to $U_i$, we need verifiable ``zero-knowledge'' proofs in the payment transactions to show the following properties without revealing $a_i(t)$ or $P_i$:
\begin{itemize}

\item[({\tt S2.1})] {\em Non-negativity of user demands}: $a_i(t) \ge 0$ for all $t$.
s
\item[({\tt S2.2})] {\em Correctness of payment}: $P_i$ is computed correctly according to Theorem~2 for each $U_i$.

\item[({\tt S2.3})] {\em Sufficient balance of payment}: ${\tt Bal}({\tt ad}_i) \ge P_i$, where ${\tt ad}_i$ is the account address of $U_i$.

\item[({\tt S2.4})] {\em Budget balance of energy storage service}: $\sum_{i=1}^N P_i = {\tt Cost}_{\tt ess}$.

\end{itemize}
These zero-knowledge proofs will be crucial to safeguard against dishonest users in cost-sharing payments.

\item[({\tt S3})] {\bf Auditing for Virtual Net Metering}: The grid operator needs to verify the agreed energy flows from the energy storage operator to users, namely, $\Big(\big({x}^-_i(t)\big)_{t=1}^T\Big)_{i=1}^N$ and $\big({x}^-(t)\big)_{t=1}^T$. To enable auditing, the energy storage operator needs to provide a receipt for each $U_i$ to certify her corresponding schedule $\big({x}^-_i(t)\big)_{t=1}^T$, but without the knowledge of $\big({x}^-_i(t)\big)_{t=1}^T$.

\end{enumerate}
We emphasize that privacy protection is considered throughout the integrated process of scheduling, cost-sharing, payment and VNM auditing, without requiring a trusted third-party.

\subsubsection{Threat Model}
\

Any users may be dishonest, who may try to cheat by paying less to energy storage service or claim more in VNM than what they ought to. These dishonest users may collude to coordinate their actions.  We aim to ensure the privacy of honest users and the correctness of scheduling, cost-sharing and payment in the presence of an adaptive adversary who may corrupt a majority of up to $N-2$ dishonest users. The adaptive adversary model provides a stronger security guarantee than the static one, where the adversary may corrupt users at any time during the protocols rather than before the protocols. A malicious adversary is more challenging than a classical semi-honest user due to her ability of deliberately deviating from the protocols for prying into others' privacy or sabotage the protocols. In case of any dishonest actions being detected, our system will abort and notify all the users. 

Note that our system is not required to identify individual dishonest user and it is fundamentally impossible \cite{BGW99} to identify a dishonest user in multi-party computation with a majority of dishonest users. There are secure multi-party computation protocols \cite{cramer2001multiparty} that can identify a dishonest user, but requiring a majority of honest users and considerable computational overhead. On the other hand, we can impose further measures to mitigate dishonesty. For example, requiring proper user authentication to prevent shilling. Or, we can require each user to pay a deposit in advance, which will be forfeited if any dishonesty is detected.

\section{Cryptographic Components}  \label{sec:crypto}

Our privacy-preserving solution relies on several basic components from cryptography. We briefly explain them in this section. More details can be found in a standard cryptography textbook (e.g., \cite{crytobk}). 

Denote by ${\mathbb Z}_p = \{0, ..., p-1\}$ the set of integers modulo $p$, for encrypting private data. For brevity, we simply write ``$x+y$'' and ``$x\cdot y$'' for modular arithmetic without explicitly mentioning``${\tt mod\ } p$''. We consider a usual finite group ${\mathbb G}$ of order $p$. We pick $g, h$ as two generators of ${\mathbb G}$, such that they can generate every element in ${\mathbb G}$ by taking proper powers, namely, for each $e \in {\mathbb G}$, there exist $x,y \in {\mathbb Z}_p$ such that $e = g^x = h^y$. The classical discrete logarithmic assumption states that given $g^x$, it is computationally hard to obtain $x$, which underlies the security of many cryptosystems.

\subsection{Cryptographic Commitments}

A cryptographic commitment allows a user to hide a secret (e.g., to hide the balances and transactions on a blockchain). We use Pedersen commitment, which is perfectly hiding (i.e., a computionally unbounded adversary cannot unlock the secret) and computationally binding (i.e., an adversary cannot associate with another secret in polynomial time). To commit secret value $x\in{\mathbb Z}_p$, a user first picks a random number ${\tt r} \in {\mathbb Z}_p$ to mask the commitment. Then, the user computes the commitment by: 
\begin{equation}
\label{eqn:pedersen}
{\tt Cm}(x, {\tt r}) = g^x \cdot h^{\tt r} \  (\mbox{mod\ } p)
\end{equation}
where $g$ is a generator of a multiplicative group ${\mathbb Z}_p^*$, $h=g^k \  (\mbox{mod\ } p)$, $k$ is a secret value and $p$ is a large prime number. 

Note that Pedersen commitment satisfies homomorphic property: ${\tt Cm}(x_1+x_2, {\tt r}_1+{\tt r}_2) = {\tt Cm}(x_1, {\tt r}_1) \cdot {\tt Cm}(x_2, {\tt r}_2)$. Sometimes, we simply write ${\tt Cm}(x)$ without specifying random ${\tt r}$. Next, we use $\Sigma$-protocol to construct zero-knowledge proofs for several useful properties of cryptographic commitments.

\subsection{Zero-knowledge Proofs (ZKP)}

In a zero-knowledge proof (of knowledge), a prover convinces a verifier of the knowledge of a secret without revealing the secret. For example, to show the knowledge of $(x, {\tt r})$ for ${\tt Cm}(x, {\tt r})$ without revealing $(x, {\tt r})$. A zero-knowledge proof of knowledge should satisfy completeness (i.e., the prover always can convince the verifier if knowing the secret), soundness (i.e., the prover cannot convince a verifier if not knowing the secret) and zero-knowledge (i.e., the verifier cannot learn the secret).

\subsubsection{$\Sigma$-Protocol}
\

$\Sigma$-Protocol is a general approach to construct zero-knowledge proofs. Given a computationally non-invertible function $f(\cdot)$ that satisfies homomorphic property $f(a+b) = f(a) + f(b)$ and $f(x)=y$, one can prove the knowledge of the concealed $x$:

\begin{enumerate}

\item First, the prover sends a commitment $y' = f(x')$, for a random $x'$, to the verifier. 

\item Next, the verifier replies with a random challenge $\beta$.

\item The prover replies with $z = x' + \beta \cdot x$ \mbox{(which does not reveal $x$)}. 

\item Finally, the verifier checks whether $f(z) \overset{?}{=} y' + \beta \cdot y$. 

\end{enumerate}

\subsubsection{$\Sigma$-Protocol Based Zero-knowledge Proofs}
\

Next, we present four crucial instances of zero-knowledge proofs based on $\Sigma$-protocol:

\begin{itemize}

\item {\bf ZKP of Commitment ({\tt zkpCm})}: Given ${\tt Cm}(x, r)$, a prover can convince a verifier of the knowledge of $x$ without revealing $(x, r)$. Denote the corresponding zero-knowledge proof by ${\tt zkpCm}[x]$.

\item {\bf ZKP of Summation ({\tt zkpSum})}: Given a set of commitments $\big({\tt Cm}(x_i, {\tt r}_i)\big)_{i=1}^n$ and $y$, a prover can convince a verifier of the knowledge of $y = \sum_{i=1}^n x_i$ without revealing $(x_i)_{i=1}^n$. Denote the corresponding zero-knowledge proof by ${\tt zkpSum}[y, (x_i)_{i=1}^n]$.

\item {\bf ZKP of Membership ({\tt zkpMbs})}: Given a set ${\mathcal X} = \{x_1, ..., x_n \}$ and ${\tt Cm}(x, {\tt r})$, a prover can convince a verifier of the knowledge of $x \in {\mathcal X}$ without revealing $x$. Denote the corresponding zero-knowledge proof by ${\tt zkpMbs}[x, {\mathcal X}]$.

\item {\bf ZKP of Non-Negativity ({\tt zkpNN})}: Given ${\tt Cm}(x, {\tt r})$, a prover can convince a verifier of the knowledge of $x \ge 0$ without revealing $x$. Denote the corresponding zero-knowledge proof by ${\tt zkpNN}[x]$.

\end{itemize}
The detailed constructions of these zero-knowledge proofs can be found in Appendix.~\ref{sec:append2}.

\subsection{Non-interactive Zero-knowledge Proofs}

An interactive zero-knowledge proof that requires  a verifier-provided challenge can be converted to a non-interactive one by Fiat-Shamir heuristic to remove the verifier-provided challenge. 

Let ${\mathcal H}(\cdot)\mapsto {\mathbb Z}_p$ be a cryptographic hash function. Given a list of commitments (${\tt Cm}_1, ..., {\tt Cm}_r$), one can map to a single hash value by  ${\mathcal H}({\tt Cm}_1|...|{\tt Cm}_r)$, where the input is the concatenated string of (${\tt Cm}_1, ..., {\tt Cm}_r$). In a $\Sigma$-protocol, one can set the challenge by $\beta ={\mathcal H}({\tt Cm}_1|$$...|{\tt Cm}_r)$, where (${\tt Cm}_1, ..., {\tt Cm}_r$) are all the commitments generated by the prover prior to the step of verifier-provided challenge (Step 2 of $\Sigma$-protocol). Hence, the prover does not wait for the verifier-provided random challenge, and instead generates the random challenge himself. The verifier will generate the same challenge following the same procedure for verification. We denote the non-interactive versions of the previous zero-knowledge proofs by {\tt nzkpCm}, {\tt nzkpSum}, {\tt nzkpMbs}, {\tt nzkpNN}, respectively. 

\subsection{Public-Private Key Signatures}

Cryptographic signatures are a standard tool to verify the authenticity of some given data. Suppose that a signer has a pair of public and private keys $(K^{\tt p}, K^{\tt s})$ for an asymmetric key cryptosystem (e.g., RSA). To sign a message $m$, the signer first maps $m$ by a cryptographic hash function ${\mathcal H}(m)$ (e.g., SHA-3). Then the signature of $m$ is the encryption ${\tt sign}_{K^{\tt s}}[m] = {\tt Enc}_{K^{\tt s}}[{\mathcal H}(m)]$. Given $(m, K^{\tt p})$, anyone can verify the signature ${\tt sign}_{K^{\tt s}}[m]$ by checking whether the decryption ${\tt Dec}_{K^{\tt p}}[{\tt sign}_{K^{\tt s}}[m]]  \overset{?}{=} {\mathcal H}(m)$.

\section{Multi-party Computation Protocol}  \label{sec:spdz}

Our privacy-preserving solution also relies on a multi-party computation protocol called SPDZ \cite{cramer2015secure, dklpss13spdz}, which allows multiple parties to jointly compute a function while concealing the private inputs. SPDZ can safeguard against a majority of dishonest users (i.e., all but one party can be dishonest), and does not require a trusted dealer for setup. For clarity, this section presents a simplified version of SPDZ. Readers can refer to \cite{cramer2015secure, dklpss13spdz} for the detailed description.

\subsection{Information-theoretical Secret Sharing} 

SPDZ relies on the notion of information-theoretical secret-sharing, whereby private data will be distributed to multiple parties, such that each party only knows a share of the data, without complete knowledge of other shares. Hence, computation of individual shares of data will not reveal the original data, unless all shares are revealed for output or verification. Several distributed computation operations can be performed locally via SPDZ, while preserving the secret sharing property.

\begin{figure}[t!]
\centering
\includegraphics[width=0.4\textwidth]{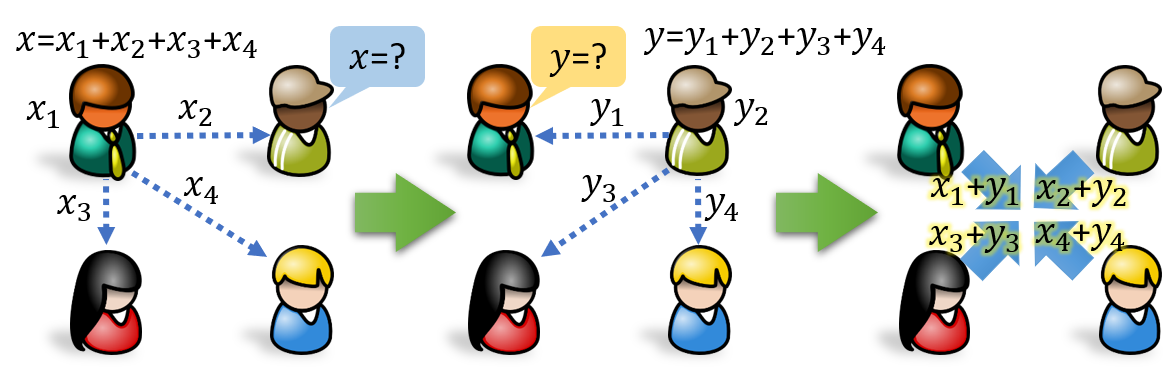}
\caption{An example of information-theoretical secret-sharing. Let us compute $x+y$, without revealing $x$ or $y$. Party 1 splits $x$ into $(x_i)_{i=1}^4$ and secretly shares each with one party. So is party 2 for $y$. Next, each party $i$ computes $z_i = x_i + y_i$ locally. Then, we can obtain the sum $x+y = \sum_{i=1}^4 z_i$, when $(z_i)_{i=1}^4$ are revealed to everyone, which however does not leak either $x_i$ or $y_i$.}
\label{fig:spdz}
\end{figure}

We consider the computation of a function of an arithmetic circuit consisting of only additions and multiplications. Suppose a private number $x$ is distributed to $n$ parties, such that each party $i$ knows a share $x_i$ only, where $x = \sum_{i=1}^n x_i$, but not knowing other shares $x_j$, where $j \ne i$. Note that a party is unable to construct $x$, without knowing all the shares. In the following, we write $\langle x \rangle$ as a {\em secretly shared} number, meaning that there is a vector $(x_1, ..., x_n)$, such that each party $i$ knows only $x_i$. Given secretly shared $\langle x \rangle$ and $\langle y \rangle$, and a public known constant $c$, the following operations can be attained by local computation at each party, and then the outcome can be assembled from the individual shares: 
\begin{enumerate}

\item[{\tt A1})] $\langle x \rangle + \langle y \rangle$ can be computed by $(x_1 + y_1, ..., x_n + y_n)$.

\item[{\tt A2})] $c \cdot \langle x \rangle$ can be computed by $(c \cdot x_1, ..., c \cdot x_n)$.

\item[{\tt A3})] $c + \langle x \rangle$ can be computed by $(c + x_1, x_2, ..., x_n)$.

\end{enumerate}
To reveal $\langle x \rangle$, each party $i$ broadcasts $x_i$ to other parties. Then each party can reconstruct $x = \sum_{i=1}^n x_i$. See an illustration in Figure~\ref{fig:spdz}.

Multiplications can also be computed by SPDZ, and the detailed description can be found in Appendix.~\ref{sec:append4}. With additions and multiplications, one can construct a large class of computation functions (including comparison and branching conditions).

However, some parties may be dishonest, who may not perform the correct local computation. To safeguard against dishonest parties, an information-theoretical message authentication code (MAC) can be used for verification. Every secretly shared number is encoded by a MAC as $\gamma(x)$, which is also secretly shared as $\langle \gamma(x) \rangle$. The basic idea is that if a dishonest party wants to modify her share $x_i$, then he also needs to modify $\gamma(x)_i$ consistently. This allows dishonesty to be detectable by checking the corresponding MAC in the final output. The detailed description of MAC can be found in Appendix.~\ref{sec:append4}. In the following, we write $\llangle  x \rrangle$ meaning that both $\langle  x \rangle$ and the respective MAC $\langle \gamma(x) \rangle$ are secretly shared among users.

\subsection{Overview of SPDZ Protocol}

The SPDZ consists of three phases, as outlined as follows:

\begin{enumerate}

\item {\em Pre-processing Phase}: In this phase, a collection of shared random numbers will be constructed to mask the private input numbers. For each private input number of party $i$, there needs a shared random number $\llangle r^i \rrangle$, where $r^i$ is revealed to party $i$ only, but not to other parties.

\item {\em Online Phase}: To secretly shares a private input number $x^i$ using $\llangle r^i \rrangle$, without revealing $x^i$, it proceeds as follows:
\begin{enumerate}

\item[1)] Party $i$ computes and reveals $z^i = x^i - r^i$ to all parties.

\item[2)] Every party sets $\llangle x^i \rrangle \leftarrow z^i + \llangle r^i \rrangle$ (see {\tt A3}).

\end{enumerate}
\smallskip

Any computation circuit with additions or multiplications can be computed by local computations (e.g., {\tt A1}-{\tt A3}). The MACs are updated accordingly to preserve the consistency. 

\item {\em Output and Validation Phase}: All MACs will be revealed for validation. If there is any inconsistency in MACs, then abort.

\end{enumerate}

The details of SPDZ protocol can be found in Appendix.~\ref{sec:append3}.

\section{Privacy-Preserving Solution for Energy Storage Sharing} \label{sec:algo}

This section presents an integrated solution for privacy-preserving energy sharing, based on blockchain, zero-knowledge proofs and multi-party computation protocol SPDZ.

\subsection{Privacy-Preserving Ledger}

First, we incorporate privacy protection to hide the transaction records on the ledger, while still allowing proper verifications for cost-sharing and VNM. As in other privacy-preserving blockchain platforms (e.g., Zether \cite{bunz20zether}), we conceal the balances and transaction values in the ledger by the respective cryptographic commitments instead of plaintext values. The accounts in the ledger will become as follows:

\begin{table}[!h]
\centering
\caption{Accounts in the privacy-preserving ledger} 
\begin{tabularx}{0.42\linewidth}{c|c|c}
    \hline  \hline
	${\tt ad}_i$ & $K^{\tt p}_i$ &${\tt Cm}({\tt Bal}({\tt ad}_i))$ \\
	\hline
	${\tt ad}_j$ & $K^{\tt p}_j$ &${\tt Cm}({\tt Bal}({\tt ad}_j))$ \\
	\hline
	... & ... & ... \\
	\hline  \hline
\end{tabularx}
\end{table}		
A multi-transaction will be concealed as ${\tt mtx}=({\tt ad}_i, {\tt ad}_{i'}, {\tt Cm}({\tt val}_i))_{i=1}^N$.

Because of concealed balances and transaction values, each user must provide ${\tt nzkpNN}[{\tt Bal}({\tt ad}_i) - {\tt val}_i]$ along with each transaction request to prove the non-negativity of the resultant balance. Otherwise, the transaction request will be denied by the ledger without the correct ${\tt nzkpNN}$.

\subsection{Privacy-Preserving Protocol $\Pi_{\tt pess}$}

We design a protocol, denoted by $\Pi_{\tt pess}$, to coordinate the users for privacy-preserving energy storage service scheduling, cost-sharing, payment and VNM settlement. We denote the energy storage operator and grid operator by ${\sf Opr}_{\sf es}$ and ${\sf Opr}_{\sf gr}$ respectively. 

Before presenting the details of the protocol, we first outline some high-level ideas:
\begin{enumerate}

\item First, the users need to secretly share private individual demands $\llangle  a_i(t) \rrangle$. Then, they can  compute aggregate demand $\llangle  a(t) \rrangle$ via SPDZ in a privacy-preserving manner.

\item To enable subsequent verification of the payment transactions and VNM, $U_i$ also needs to announce commitment ${\tt Cm}(a_i(t))$ to each other. However, a dishonest user may use inconsistent commitment ${\tt Cm}(a_i(t))$ with respect to the secretly shared $\llangle  a_i(t) \rrangle$. To show the consistency between ${\tt Cm}(a_i(t))$ and $\llangle  a_i(t) \rrangle$, all users need to create zero-knowledge proof of commitment ${\tt zkpCm}[a_i(t)]$ via SPDZ using secretly shared $\llangle  a_i(t) \rrangle$. If ${\tt zkpCm}[a_i(t)]$ is verified to be correct, then ${\tt Cm}(a_i(t))$ and $\llangle  a_i(t) \rrangle$ are consistent. Each user also creates a zero-knowledge proof of non-negativity ${\tt nzkpNN}[a_i(t)]$.

\item After verifying ${\tt zkpCm}[a_i(t)]$ and ${\tt nzkpNN}[a_i(t)]$, the users reveal $\llangle  a(t) \rrangle$ and verify the corresponding MAC to ensure the integrity of $a(t)$. Then, the users compute the energy storage service schedule with the knowledge of $\big(a(t)\big)_{t=1}^T$.

\item Next, $U_i$ can make her payment $P_i (= P_i^{\tt pp} \mbox{\ or\ } P_i^{\tt ega})$ by Theorem~2. The users jointly compute the total payments $\sum_i^N P_i$ via SPDZ in a privacy-preserving manner to ensure that the difference between ${\tt Cost_{\tt ess}}$ and $\sum_i^N P_i$ is within a negligible rounding error $\varepsilon$, such that $|{\tt Cost_{\tt ess}}-\sum_i^N P_i| < \varepsilon$. The users agree and set ${\tt Cost_{\tt ess}}=\sum_i^N P_i$.

\item To make cost-sharing payments for energy storage service on the ledger, the users need to create a zero-knowledge proof that $\sum_{i=1}^N P_i = {\tt Cost}_{\tt ess}$ via SPDZ. $U_i$ also creates ${\tt nzkpNN}[{\tt Bal}({\tt ad}_i)-P_i]$ locally. Then, the users submit a multi-transaction request with relevant zero-knowledge proofs to the ledger.

\item After the completion of multi-transaction of payments, the energy storage service schedule is executed. Afterwards, ${\sf Opr}_{\sf es}$ signs ${\tt Cm}\big(x^-_i(t)\big)$ as a receipt on the ledger for each user. Note that ${\tt Cm}\big(x^-_i(t)\big)$ can be generated based on ${\tt Cm}\big(a_i(t)\big)$ and the energy storage service schedule.

\item The users request VNM settlement with ${\sf Opr}_{\sf gr}$, who will verify $x^-_i(t)$ from the signed ${\tt Cm}\big(x^-_i(t)\big)$ on the ledger. 

\end{enumerate}

\medskip

Next, we present the details of the privacy-preserving protocol $\Pi_{\tt pess}$, consisting of four stages (Initialization, Pre-operation Scheduling, Cost-sharing Payment \& Operation and Post-operation VNM Settlement), as follows:

\subsubsection*{\bf Stage 0: Initialization}
\

In this stage, the system parameters are chosen and the pre-processing phase of SPDZ is executed among the users. See Appendix.~\ref{sec:append3} for detailed SPDZ pre-processing phase.

\medskip

\begin{longfbox}[border-break-style=none,border-color=\#bbbbbb,background-color=\#eeeeee,breakable=true,,width=\linewidth]
{\em Initialization}: 
\begin{enumerate}

\item Choose and announce a multiplicative group ${\mathbb Z}_p^*$, two generators $g, h \in {\mathbb Z}_p^*$ and hash function ${\mathcal H}(\cdot)\mapsto {\mathbb Z}_p$ as public information to all users. Note that $g$ and $h$ can be obtained via a coin-tossing protocol \cite{cointossing} among the users such that $\log_{g} h$ is unknown due to the hardness of discrete logarithm.

\item The energy storage operator announces ${\tt p}_{\tt s}, {\tt e}_{\tt c}, {\tt e}_{\tt d}, {\tt r}_{\tt c}, {\tt r}_{\tt d},$ $\big(p(t),$\\ ${\tt B}(t)\big)_{t=1}^T$ as public information to all users.

\item Initialize SPDZ pre-processing phase among all users.

\end{enumerate}
\end{longfbox}

\medskip

\subsubsection*{\bf Stage 1: Pre-operation Scheduling}
\

In this stage, the users will compute their aggregate day-ahead demands via SPDZ. The users also need to make commitments of their individual demands $\big(a_i(t)\big)_{t=1}^T$, which will be used for auditing in VNM. We ensure that the individual demands shared via SPDZ match the ones being committed. This can be accomplished by computing zero-knowledge proof of commitment ${\tt zkpCm}[a_i(t)]$ using the secretly shared $\llangle a_i(t)\rrangle$. Next, the users will compute the optimal energy storage service schedule in $({\tt P2})$ based on aggregate demands $\big(a(t)\big)_{t=1}^T$. 

\medskip

\begin{longfbox}[border-break-style=none,border-color=\#bbbbbb,background-color=\#eeeeee,breakable=true,,width=\linewidth]
{\em Protocol $\Pi_{\tt pess}^{(1)}$}: 
\begin{enumerate}

\item $U_i$ commits $C_i(t)={\tt Cm}\big(a_i(t), {\tt r}_i(t)\big)$ for all $t$ and announces $\big(C_i(t)\big)_{t=1}^T$ to all users with $\big({\tt nzkpNN}[a_i(t)]\big)_{t=1}^T$, where ${\tt r}_i(t)$ is a random masking number. All users verify ${\tt nzkpNN}[a_i(t)]$. If verification of ${\tt nzkpNN}[a_i(t)]$ fails, announce {\tt Abort}.

\item $U_i$ secretly shares $\llangle a_i(t)\rrangle$ and $\llangle {\tt r}_i(t)\rrangle$ via SPDZ for all $t$.

\item To show the equality of $a_i(t)$ in $\llangle a_i(t)\rrangle$ and $C_i(t)$, $U_i$ constructs an ${\tt zkpCm}[a_i(t)]$ distributedly via SPDZ:

\begin{enumerate}

\item $U_i$ randomly generates $\big(a'_{i}(t), {\tt r}'_{i}(t)\big) \in {\mathbb Z}_p$ and secretly shares as $\llangle a'_i(t)\rrangle$ and $\llangle {\tt r}'_i(t)\rrangle$. Then $U_i$ announces $C'_i(t) = {\tt Cm}\big(a'_{i}(t), {\tt r}'_{i}(t)\big)$ to all users. Note that all the users must complete this step before proceeding to the next step to produce a common random challenge $\beta(t)$.

\item All the users conduct a coin-tossing protocol to obtain a random challenge $\beta(t)$. Firstly, each user announces a commitment $C''_i(t)$ of a randomly generated number ${\tt r}''_i(t) \in {\mathbb Z}_p$. Then all the users reveal ${\tt r}''_i(t)$ and compute a random challenge $\beta(t) = \sum_{i=1}^N {\tt r}''_i(t)$. 

\item All users compute and reveal $\llangle z_{a_i(t)} \rrangle = \llangle a'_{i}(t) \rrangle + \beta(t) \cdot \llangle a_i(t)\rrangle$ and $\llangle z_{{\tt r}_i(t)} \rrangle = \llangle {\tt r}'_{i}(t) \rrangle + \beta(t) \cdot \llangle {\tt r}_i(t) \rrangle$ for $U_i$. 

\item This creates ${\tt zkpCm}[a_i(t)] = \{C'_i(t) , z_{a_i(t)},  z_{{\tt r}_i(t)} \}$. All users verify ${\tt zkpCm}$ by checking
\[
g^{z_{a_i(t)}} \cdot h^{z_{{\tt r}_i(t)}} \overset{?}{=}  C'_i(t)  \cdot {\tt Cm}\big(a_i(t), {\tt r}_i(t)\big)^{\beta(t)}
\]

\item If the verification of ${\tt zkpCm}$ fails, announce {\tt Abort}.

\end{enumerate}

\item The users compute $\llangle a(t) \rrangle  \leftarrow \sum_{i=1}^N\llangle a_i(t)\rrangle $ for all $t$ via SPDZ. Reveal  $\llangle a(t) \rrangle$ to all users. Check MAC of $\llangle a(t) \rrangle$. If the MAC check fails, announce {\tt Abort}.

\item The users solve $({\tt P2})$ using $\big({a}(t)\big)_{t=1}^T$ for $\big({x}^+(t), {x}^-(t), {y}(t)\big)_{t=1}^T$

\item All users compute ${\tt Cost}_{\tt ess} \triangleq \sum_{t=1}^T \big( p(t) + {\tt p}_{\tt s} \big) \cdot {x}^+(t)$ and  ${\tt Cost}_{\tt org} \triangleq \sum_{t=1}^T  p(t) \cdot {x}^-(t)$,\quad  $\hat{\tt p}(t) \triangleq \frac{{x}^-(t) \cdot p(t)}{a(t)}$.

\end{enumerate}

\end{longfbox} 

\medskip

\subsubsection*{\bf Stage 2: Cost-sharing Payment \& Operation}
\

In this stage, the users will split the cost of energy storage service based on proportional or egalitarian cost-sharing scheme via SPDZ. The users compute the payment commitments and verify the validity of ${\tt Cost}_{\tt ess}$ by comparing with $\sum_{i=1}^N P_i$. Before issuing the multi-transaction, the users compute ${\tt nzkpSum}[{\tt Cost}_{\tt ess}, (P_i )_{i=1}^N]$ to satisfy Theorem~3. The users then make energy storage service payments via privacy-preserving blockchain. After receiving the payments, the energy storage operator will issue verifiable receipts on the ledger. 

\medskip

\begin{longfbox}[border-break-style=none,border-color=\#bbbbbb,background-color=\#eeeeee,breakable=true,,width=\linewidth]
{\em Protocol $\Pi_{\tt pess}^{(2)}$}: 

\begin{enumerate}

\item $U_i$ computes $P_i = P_i^{\tt pp}$ (or $P_i^{\tt ega}$) by Theorem~2, and announces commitment ${\tt Cm}(P_i, {\tt r}_i)$ to all users, and secretly shares $\llangle {\tt r}_i \rrangle$ via SPDZ.

\item The users also compute the total payments via SPDZ by
\begin{equation} \sum_{i=1}^N P_i =
\begin{cases}
\displaystyle \sum_{i=1}^N P_i^{\tt pp}   =  \sum_{t=1}^T \big(\frac{{\tt Cost}_{\tt ess} \cdot \hat{\tt p}(t)}{{\tt Cost}_{\tt org}} \cdot \sum_{i=1}^N \llangle a_i(t)\rrangle \big), \\  
\displaystyle \sum_{i=1}^N  P_i^{\tt ega}  =  \sum_{t=1}^T \Big(\sum_{i=1}^N \big(\llangle a_i(t)\rrangle \cdot \hat{\tt p}(t)  - \frac{{\tt Cost}_{\tt org}  - {\tt Cost}_{\tt ess}}{N}\big) \Big)
\end{cases}
\end{equation}
and check if satisfying $|{\tt Cost}_{\tt ess} -\sum_{i=1}^N P_i | < \varepsilon$, where $\varepsilon$ is a small fault-tolerant factor, which restricts the rounding error, arising from computing $\frac{{\tt Cost}_{\tt ess}}{{\tt Cost}_{\tt org}}$ or $\hat{\tt p}(t)$. If satisfied, the users let ${\tt Cost}_{\tt ess}\leftarrow\sum_{i=1}^N P_i$. Otherwise, announce {\tt Abort}.

\item The users compute ${\tt nzkpSum}[{\tt Cost}_{\tt ess}, (P_i )_{i=1}^N]$ via SPDZ distributely:

\begin{enumerate}

\item $U_i$ randomly generates and secretly shares $\llangle {\tt r}'_i \rrangle \in {\mathbb Z}_p$ before announcing 
${\tt Cm}(0, {\tt r}'_{i})$.

\item All users compute $C' = \prod_{i=1}^N {\tt Cm}(0, {\tt r}'_{i})$ and obtain a random challenge $\beta = {\mathcal H}(C')$. Then all users compute $\llangle z_{\tt r} \rrangle = \sum_{i=1}^n \llangle {\tt r}'_i \rrangle + \beta \cdot \sum_{i=1}^n \llangle {\tt r}_i \rrangle$. Then reveal $\llangle z_{\tt r} \rrangle$.

\item This creates ${\tt nzkpSum}[{\tt Cost}_{\tt ess}, (P_i )_{i=1}^N] = \{C' , z_{{\tt r}} \}$. All users verify ${\tt nzkpSum}$ by checking:
\[
g^{\beta \cdot {\tt Cost}_{\tt ess}}\cdot h^{z_{\tt r}} \overset{?}{=}  C'\cdot\prod_{i=1}^n {\tt Cm}(P_i, {\tt r}_i)^\beta
\]
If the verification of ${\tt nzkpSum}$ fails, announce {\tt Abort}.

\end{enumerate}

\item $U_i$ computes ${\tt nzkpNN}[{\tt Bal}({\tt ad}_i) - P_i]$ based on ${\tt Cm}(P_i)$.

\item Let the account address of ${\sf Opr}_{\sf es}$ be ${\tt ad}_{\tt ess}$. The users submit a multi-transaction request 
\[
{\tt mtx} = \big({\tt ad}_i, {\tt ad}_{\tt ess}, {\tt Cm}(P_i)\big)_{i=1}^N
\]
to the ledger, along with 
\[
\qquad \ \ {\tt nzkpSum}[{\tt Cost}_{\tt ess}, (P_i )_{i=1}^N] \mbox{\ and\ } {\tt nzkpNN}[{\tt Bal}({\tt ad}_i) - P_i]_{i=1}^N
\]

\item The ledger verifies ${\tt nzkpSum}$ and ${\tt nzkpNN}$ before proceeding the transaction. If the verification fails, announce {\tt Abort}. 
 
 \item The users provide the schedule $\big({a}(t), {x}^+(t), {x}^-(t), {y}(t)\big)_{t=1}^T$ to ${\sf Opr}_{\sf es}$.  After the transaction completes, ${\sf Opr}_{\sf es}$ will execute the schedule.

\end{enumerate}

\end{longfbox}

\medskip

\subsubsection*{\bf Stage 3:  Post-operation VNM Settlement}
\

In this stage, ${\sf Opr}_{\sf es}$ will sign the receipts of individual energy storage service $\big(x^-_i(t)\big)_{t=1}^T$. The receipts will be stored on the ledger. When the users request VNM settlement with ${\sf Opr}_{\sf gr}$, ${\sf Opr}_{\sf gr}$ will verify their claims by the receipts on the ledger.

\medskip

\begin{longfbox}[border-break-style=none,border-color=\#bbbbbb,background-color=\#eeeeee,breakable=true,,width=\linewidth]
{\em Protocol $\Pi_{\tt pess}^{(3)}$}: 

\begin{enumerate}

\item The users upload $\big({\tt Cm}(a_i(t))_{t=1}^T\big)_{i=1}^N$ (from Stage 1) to the ledger. 

 \item ${\sf Opr}_{\sf es}$ computes commitment ${\tt Cm}(x^-_i(t))$ $= {\tt Cm}(a_i(t))^{\frac{x^-(t)}{a(t)}}$ (according to Theorem~2) for all $t$ and $i$.  
 
 \item Let the public-private keys of ${\sf Opr}_{\sf es}$ be $(K^{\tt p}_{\tt ess}, K^{\tt s}_{\tt ess})$. ${\sf Opr}_{\sf es}$ signs ${\tt sign}_{K^{\tt s}_{\tt ess}}[$ ${\tt Cm}(x^-_i(t)) ]$ along with ${\tt Cm}(x^-_i(t))$ to be stored on the ledger.
 
\item ${\sf Opr}_{\sf es}$ prepares VNM and provides energy export profile $\big(x^-(t)\big)_{t=1}^T$ to ${\sf Opr}_{\sf gr}$.

\item $U_i$ submits a claim for reimbursement by referring to receipt ${\tt Cm}(x^-_i(t))$ and ${\tt sign}_{K^{\tt s}_{\tt ess}}[{\tt Cm}(x^-_i(t)) ]$ on the ledger. $U_i$ also reveals $x^-_i(t)$ to ${\sf Opr}_{\sf gr}$ to prove the validity.

\item ${\sf Opr}_{\sf gr}$ verifies ${\tt sign}_{K^{\tt s}_{\tt ess}}[{\tt Cm}(x^-_i(t)) ]$ by public key $K^{\tt p}_{\tt ess}$, and compares the energy demand profile $\big(a_i(t)\big)_{t=1}^T$ with $\big(x^-_i(t)\big)_{t=1}^T$. If the verification is consistent, ${\sf Opr}_{\sf gr}$ will deduct the amount $\sum_{t=1}^T x^-_i(t) \cdot p(t)$ from $U_i$'s total payment. 

\end{enumerate}

\end{longfbox}

\medskip

{\bf Remarks:} In protocol $\Pi_{\tt pess}$, each user is required to input her privacy demand $\big(a(t)\big)_{t=1}^T$ in two privacy-preserving ways: (1) commitment ${\tt Cm}(a_i(t))$, and (2) secretly shared value in SPDZ $\llangle a_i(t)\rrangle$. While the commitment ${\tt Cm}(a_i(t))$ is used to generate other zero-knowledge proofs for payments and VNM, the secretly shared $\llangle a_i(t)\rrangle$ is used to compute service scheduling and cost-sharing. Both inputs should be consistent (i.e., checked by ${\tt zkpCm}[a_i(t)]$ that is constructed via SPDZ). Also, ${\tt Cm}(a_i(t))$ can be used to construct ${\tt Cm}(P_i)$ and ${\tt Cm}(x^-_i(t))$ without the knowledge of $a_i(t)$, because of its homomorphic property for these constructions.

See Appendix.~\ref{sec:append4} for the security analysis of $\Pi_{\tt pess}$ for satisfying security requirements {\tt S1}-{\tt S3}.

\begin{figure*}[t]
\centering  
\includegraphics[width=\linewidth]{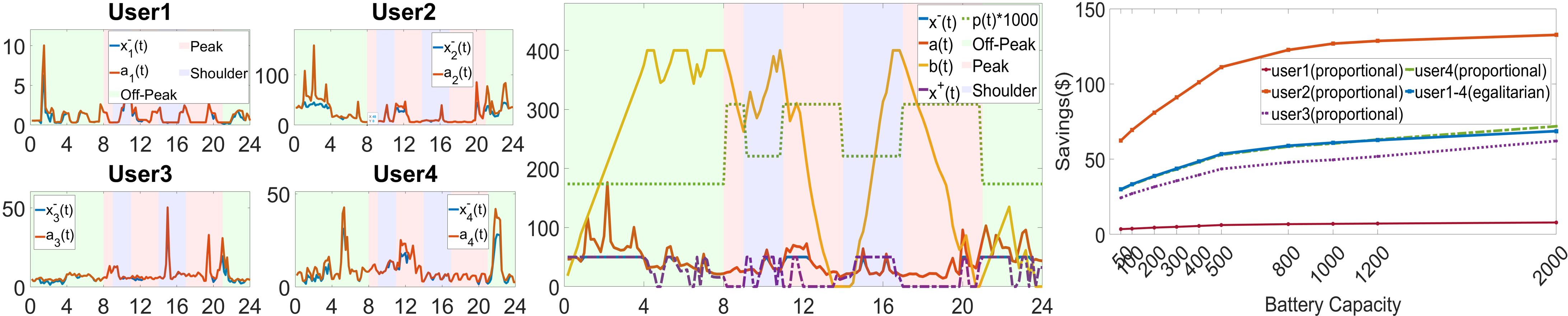} \vspace{-20pt}
\caption{Data trace of energy storage service schedule.} \label{fig:exp1} 
\end{figure*}

\begin{figure*}[t!]
\centering
\begin{subfigure}[t]{0.27\textwidth}
\includegraphics[width=1\linewidth]{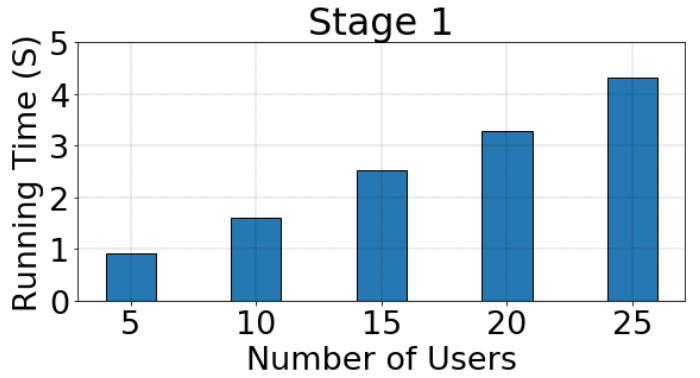}
\caption{Computational overhead}
\label{fig:computational}
\end{subfigure}  
\begin{subfigure}[t]{0.46\textwidth}
\includegraphics[width=1\linewidth]{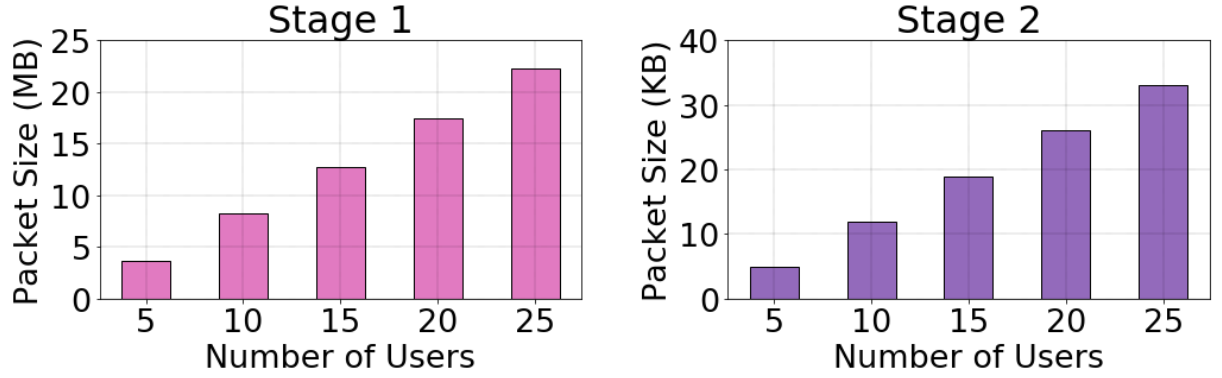}
\caption{Communication overhead}
\label{fig:communication}
\end{subfigure} 
\begin{subfigure}[t]{0.25\textwidth}
\includegraphics[width=1\linewidth]{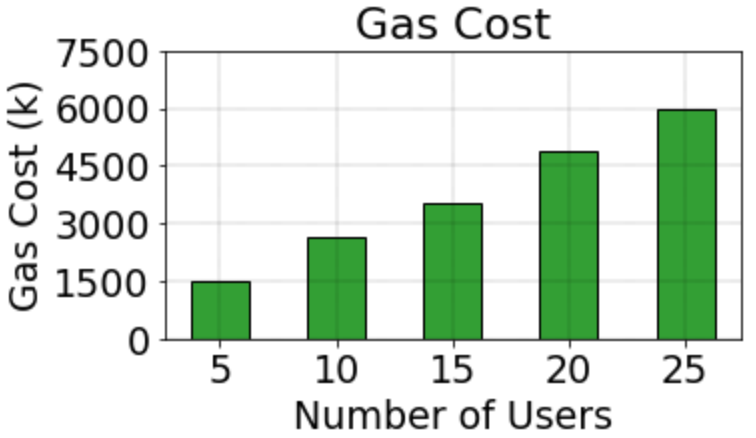}
\caption{Gas cost of {\tt executeTransaction()}}
\label{fig:gascost}
\end{subfigure}  \vspace{-15pt}
\caption{System performance and smart contract gas costs.}
\end{figure*}

\section{Evaluation} \label{sec:eval}

In this section, we present an evaluation study of our solution, including the effectiveness of energy storage sharing, multi-party computation protocol performance and the incurred cost of smart contract implementation on a practical blockchain platform.

\subsection{Energy Storage Service Scheduling}

We first evaluate the effectiveness of energy storage sharing. We selected 120 users from the Smart* microgrid dataset \cite{data34}. We consider a single 24-hour period, from midnight to next midnight. %We consider the parameters as: charging and discharging efficiency ratios ${\tt e}_{\tt c} = 1$, ${\tt e}_{\tt d} = 1$,  and energy storage service fee ${\tt p}_{\tt s}$ = \$0.05/kWh. We consider 4 users. The capacity of energy storage  ${\tt B}$ is scaled from 50 kWh to 2000 kWh, and charging and discharging rate constraints from 50 kW/min to 1000 kW/min. The pricing plan is given as follows: \$0.3088/kWh (peak rate for 7am-10am, 12pm-3pm, 6pm-10pm), \$0.2208/kWh (shoulder rate for 10am-12pm, 3pm-6pm), \$0.1738/kWh (off-peak rate for 10pm-6am).

In particular, we present the temporal data trace of scheduled energy storage services for 4 users. We observe that each of the user can utilize energy storage discharging during peak-hour. Most users acquire energy from the grid during off-peak-hour, and partially during shoulder-hour. Next, we study the saving ($\Delta_i$) of each user. In Figure~\ref{fig:exp1}, we study different cost-sharing schemes of energy storage service for 4 users. We observe that proportional cost-sharing gives each user at the same percentage of saving of 28.78\% when energy storage $B$ is 400 kWh, whereas different user has different percentage of saving with egalitarian cost-sharing, but with the same amount of saving of \$37.90. 

\subsection{SPDZ Performance} 

Next, we evaluate the performance of SPDZ in stages 1-2 of $\Pi_{\text{pess}}$. We skip stage 3 due to its negligible performance compared with stages 1-2. We consider 144 time slots in a single 24-hour period. All the results were averaged over 20 instances.

\subsubsection{Computational Overhead}

We scaled the number of users from 5 to 25. Figure~\ref{fig:computational} displays the average running time incurred at each user in the stage 1. The running time shows a linearly growing trend with the increased number of users. The running time starts from about 0.91 seconds with 5 users to around 4.32 seconds with 25 users. We skip the displaying of the stage 2 due to the negligible computational overhead of only several milliseconds. 

\subsubsection{Communication Overhead}

Figure~\ref{fig:communication} shows the average total volume of the transmission data in stages 1 and 2. It is evident that the total transmission amount scale linearly with the growing number of users in both stages. The data volume increases from 3.62 MB with 5 users to about 26.89 MB with 25 users in stage 1. In contrast, the data volume in stage 2 starts from a merely 5 KB with 5 users to about 40 KB with 25 users. Thus, the total data volume in stage 1 dominates the entire protocol. % for the reason that each user has to deal with the data of 144 timeslots, whereas it suffices to only handle the total payment in stage 2.
  
\subsection{Ethereum Smart Contract Gas Costs}  

We implemented the payment systems as a smart contract on real-world Ethereum blockchain platform. The smart contract is specified by Solidity programming language \cite{solidity}. We outline some implementation components as follows:

\begin{enumerate}
\item{\em Pedersen.} This component aims to realize the underlying Pedersen commitment scheme.

\item{\em ESToken.} We created a Ethereum-based cryptocurrency {\em ESToken} for our energy sharing scenario, whereby users are able to pay the energy cost without revealing the true payments. 

\item{\em MultiSignature.} This component allows the users to submit a multi-transaction request, where the transaction will proceed unless all the involved users validate the transaction. There are three main methods: {\tt submitTransaction()}, \\ {\tt confirmTransaction()} and {\tt executeTransaction()}.

\end{enumerate}

Distributed miners will execute the compiled bytecode of the smart contracts in {\em Ethereum Virtual Machine}. Miners will charge additional Ether/ETH (Ethereum native crytocurrency) called gas costs, because the extra computational tasks incurred by smart contracts will be broadcast throughout the blockchain. Gas costs are used to measure the amount of computational resources to execute the operations required by a transaction. We measured the incurred gas costs by our smart contracts and used a 255-bit prime number $q$ for Pedersen commitment since Solidity supports at most 256-bit numbers. We employed Truffle Suite \cite{truffle} as the Ethereum development framework to test and measure the average gas costs\footnote{The actual gas costs may vary based on the random generated parameters of the zero-knowledge proofs. Here, we show the average gas costs.} of Multi-Signature methods, shown in Table \ref{tab:gascost}, where $N$ is the number of users and $N_b$ indicates the number of bits to represent the plaintext payment in the ${\tt nzkpMbs}$. A transaction initiator must pay sufficient amount of gas costs to the miner, who creates transaction blocks on the network. The {\em gasPrice} in a transaction allows the transaction initiator to set the gas price that she is willing to pay. The higher the gas prices, the higher probability the transaction will be chosen by the miner in a block. We use the standard gas price 54 Gwei and Ether price\footnote{The Ether price quote was obtained on 25 Oct 2021.} \$4068 USD/ETH to estimate the equivalent transaction cost in Ether (see \cite{ethgasstation} and \cite{coindesk}).

\begin{table}[!h]
\centering
\caption{Table of gas costs for multi-signature methods} 
\begin{tabular}{@{}c|@{ }c@{ }|@{ }c@{ }|@{ }c@{ }|c@{}} 
\hline \hline
        & Gas Cost & Ether & USD {\scriptsize(as on 25 Oct 21)} & |Input| (bytes)                 \\ \hline \hline
{\tt submit()}  & 106k   & 0.0057 &  \$ 23.2 & 64 + 64  $\times N$ \\ \hline
{\tt confirm()} & 3600k   & 0.1944 & \$ 790.8 & 128 + 180 $\times  N_b$    \\ \hline \hline
\end{tabular}
\label{tab:gascost}
\end{table}

Figure~\ref{fig:gascost} presents the gas cost of {\tt executeTransaction()} function. We observe that the gas cost is linearly proportional to the number of involved users, starting from 1437k (0.0776 ether, \$ 315.7) with 5 users to 5986k (0.3232 ether, \$ 1314.8) with 25 users, since the verification of ${\tt nzkpSum}$ depends on the number of users. Overall, we observe only moderate incurred costs by our smart contract, which are comparable to other privacy-preserving smart contract studies in the literature.
\section{Extensions and Limitations} \label{sec:extension}

In this section, we discuss some possible extensions to enhance our privacy-preserving energy storage sharing solution. We also discuss some limitations of our current solution.

\subsection{Flexible Demands and Hour-ahead Scheduling}

Our current energy storage service scheduling is designed to operate in a day-ahead manner. There is a limitation that users are not supposed to alter their requested demand (at least, for the part of demand that is allocated to be satisfied by scheduled energy storage discharging). A more flexible approach is desirable to incorporate both day-ahead and hour-ahead scheduling processes, which is called {\em dual service scheduling}. The hour-ahead scheduling process allows the users to request energy storage service in a much shorter time-window. 

The dual service scheduling can be achieved by dividing the capacity of energy storage into two parts: (1) ${\tt B}_{\tt d}(t)$ as the capacity for day-ahead service scheduling and (2) ${\tt B}_{\tt h}(t)$ for hour-ahead service scheduling, such that ${\tt B}_{\tt d}(t) + {\tt B}_{\tt h}(t) = {\tt B}(t)$. Both service scheduling processes can run separately, with hour-ahead scheduling running before each hour.  Our privacy-preserving service scheduling process can be extended straightforwardly to operate the two parts of service scheduling in tandem.

\subsection{Virtual Net Metering Service Fees} 

In our current privacy-preserving service scheduling process, we assume that the energy storage operator can transfer the credits as a result of exporting energy through discharging energy storage via virtual net metering to the users, with no additional service fees from the grid operator. Namely, the energy storage operator can always transfer the full credits of ${x}^-(t) \cdot p(t)$ to compensate the users' consumption at each timeslot $t$. 

However, the grid operator may charge additional fees to attain virtual net metering, in particular, when the energy storage operator is not located in the same local grid network as the users. For instance, the grid operator charges fees as ${\tt s}_{\tt vnm}$ percent of the credits transferred in virtual net metering. Namely, when the energy storage operator transfers the credits of ${x}^-(t) \cdot p(t)$ to the users, the users actually receives $(1-{\tt s}_{\tt vnm}) \cdot {x}^-(t) \cdot p(t)$. 

One possible approach to incorporate in our privacy-preserving service scheduling process is to modify Cons.~\raf{eqn:SoC} as
\begin{equation}
b(t+1) - b(t)  =   {\tt e}_{\tt c} {x}^+(t) -  {\tt e}_{\tt d} \tfrac{1}{(1-{\tt s}_{\tt vnm})} \Big(\sum_{i=1}^N {x}^-_i(t) \Big)
\end{equation}
That is, we reduce the actual amount of discharged energy from the energy storage as to incorporate the additional service fees for virtual net metering from the grid operator.

\subsection{Reducing Gas Cost on Ethereum}
Although our current solution incurs a moderate gas cost on Ethereum, this is still considerable when the Ether price has increased significantly in the recent years. High Ether price has deterred many real-world smart contract projects from operation. We seek to improve the gas cost of our smart contract solution. There are a number of possible improvements. First, rather than storing the receipts on the ledger, which can take up a considerable storage space and costs extra gas cost, we can store a small hash pointer instead on the ledger. The users can verify the contents of the receipts by matching with the hash pointer. Second, there should be more efficient zero-knowledge proofs that can be executed on smart contract. One possible option is Bulletproofs \cite{bulletproof18}, which are more succinct zero-knowledge proofs than $\Sigma$-protocol. Bulletproofs have been employed in certain privacy-preserving blockchain platforms (e.g. Monero, Zether \cite{monero,bunz20zether}). Third, even though we implemented our solution as a smart contract on permissionless Ethereum blockchain platform, our solution can also be implemented on a permissioned blockchain platform, on which the gas cost is not a major concern.
\section{Conclusion} \label{sec:concl}

In this paper, we provide a novel approach to support third-party energy storage sharing without compromising the privacy of individual users. In our privacy-preserving solution, an energy storage operator is only revealed the minimal information to schedule energy storage operations, without knowing users' private demands. At the same time, the users can divide the cost of energy storage service fairly among themselves without knowing each other's demands.  Our solution can effectively safeguard against a majority of dishonest users, without requiring trusted third-parties. We implemented our solution as a smart contract on Ethereum blockchain platform, which incurs moderate overhead and gas costs in practice.

In future work, we will support robustness against potentially dishonest energy storage operators. For instance, we can require an energy storage operator to prove that her energy export profile matches the service schedules in order to receive the payments from users. We will also explore the support for peer-to-peer energy storage sharing by distributing service scheduling and cost-sharing computation among the end users themselves.

\bibliographystyle{ACM-Reference-Format}
\bibliography{reference}


\begin{thebibliography}{40}

%%% the Web address.

\ifx \showCODEN    \undefined \def \showCODEN     #1{\unskip}     \fi
\ifx \showDOI      \undefined \def \showDOI       #1{#1}\fi
\ifx \showISBNx    \undefined \def \showISBNx     #1{\unskip}     \fi
\ifx \showISBNxiii \undefined \def \showISBNxiii  #1{\unskip}     \fi
\ifx \showISSN     \undefined \def \showISSN      #1{\unskip}     \fi
\ifx \showLCCN     \undefined \def \showLCCN      #1{\unskip}     \fi
\ifx \shownote     \undefined \def \shownote      #1{#1}          \fi
\ifx \showarticletitle \undefined \def \showarticletitle #1{#1}   \fi
\ifx \showURL      \undefined \def \showURL       {\relax}        \fi
% invisible to TeX
\providecommand\bibfield[2]{#2}
\providecommand\bibinfo[2]{#2}
\providecommand\natexlab[1]{#1}
\providecommand\showeprint[2][]{arXiv:#2}

\bibitem[\protect\citeauthoryear{Andoni, Robu, Flynn, Abram, Geach, Jenkins,
  McCallum, and Peacock}{Andoni et~al\mbox{.}}{2019}]%
        {a19survey}
\bibfield{author}{\bibinfo{person}{Merlinda Andoni}, \bibinfo{person}{Valentin
  Robu}, \bibinfo{person}{David Flynn}, \bibinfo{person}{Simone Abram},
  \bibinfo{person}{Dale Geach}, \bibinfo{person}{David~P. Jenkins},
  \bibinfo{person}{Peter McCallum}, {and} \bibinfo{person}{Andrew Peacock}.}
  \bibinfo{year}{2019}\natexlab{}.
\newblock \showarticletitle{Blockchain technology in the energy sector: A
  systematic review of challenges and opportunities}.
\newblock \bibinfo{journal}{\emph{Renewable and Sustainable Energy Reviews}}
  \bibinfo{volume}{100} (\bibinfo{year}{2019}), \bibinfo{pages}{143--174}.
\newblock


\bibitem[\protect\citeauthoryear{Barker, Mishra, Irwin, Cecchet, Shenoy, and
  Albrecht}{Barker et~al\mbox{.}}{2012}]%
        {data34}
\bibfield{author}{\bibinfo{person}{Sean Barker}, \bibinfo{person}{Aditya
  Mishra}, \bibinfo{person}{David Irwin}, \bibinfo{person}{Emmanuel Cecchet},
  \bibinfo{person}{Prashant Shenoy}, {and} \bibinfo{person}{Jeannie Albrecht}.}
  \bibinfo{year}{2012}\natexlab{}.
\newblock \showarticletitle{Smart*: An Open Data Set and Tools for Enabling
  Research in Sustainable Homes}. In \bibinfo{booktitle}{\emph{SustKDD}}.
\newblock


\bibitem[\protect\citeauthoryear{Ben-Or, Goldwasser, and Wigderson}{Ben-Or
  et~al\mbox{.}}{1988}]%
        {BGW99}
\bibfield{author}{\bibinfo{person}{Michael Ben-Or}, \bibinfo{person}{Shafi
  Goldwasser}, {and} \bibinfo{person}{Avi Wigderson}.}
  \bibinfo{year}{1988}\natexlab{}.
\newblock \showarticletitle{Completeness Theorems for Non-Cryptographic
  Fault-Tolerant Distributed Computation}. In \bibinfo{booktitle}{\emph{Annual
  ACM Symposium on Theory of Computing (STOC)}}.
\newblock


\bibitem[\protect\citeauthoryear{Ben-Sasson, Chiesa, Garman, Green, Miers, and
  Virza}{Ben-Sasson et~al\mbox{.}}{2014}]%
        {zcash}
\bibfield{author}{\bibinfo{person}{Eli Ben-Sasson}, \bibinfo{person}{Alessandro
  Chiesa}, \bibinfo{person}{Christina Garman}, \bibinfo{person}{Matthew Green},
  \bibinfo{person}{Ian Miers}, {and} \bibinfo{person}{Eran Tromerand~Madars
  Virza}.} \bibinfo{year}{2014}\natexlab{}.
\newblock \showarticletitle{Zerocash: Decentralized Anonymous Payments from
  Bitcoin}. In \bibinfo{booktitle}{\emph{IEEE Symposium on Security and
  Privacy}}.
\newblock


\bibitem[\protect\citeauthoryear{Buchanan}{Buchanan}{2017}]%
        {crytobk}
\bibfield{author}{\bibinfo{person}{William~J. Buchanan}.}
  \bibinfo{year}{2017}\natexlab{}.
\newblock \bibinfo{booktitle}{\emph{Cryptography}}.
\newblock \bibinfo{publisher}{River Publishers}.
\newblock


\bibitem[\protect\citeauthoryear{Bunz, Agrawal, Zamani, and Boneh}{Bunz
  et~al\mbox{.}}{2020}]%
        {bunz20zether}
\bibfield{author}{\bibinfo{person}{Benedikt Bunz}, \bibinfo{person}{Shashank
  Agrawal}, \bibinfo{person}{Mahdi Zamani}, {and} \bibinfo{person}{Dan Boneh}.}
  \bibinfo{year}{2020}\natexlab{}.
\newblock \showarticletitle{Zether: Towards Privacy in a Smart Contract World}.
  In \bibinfo{booktitle}{\emph{Financial Cryptography and Data Security (FC)}}.
\newblock


\bibitem[\protect\citeauthoryear{Bunz, Bootle, Boneh, Poelstra, Wuille, and
  Maxwell}{Bunz et~al\mbox{.}}{2018}]%
        {bulletproof18}
\bibfield{author}{\bibinfo{person}{B. Bunz}, \bibinfo{person}{J. Bootle},
  \bibinfo{person}{D. Boneh}, \bibinfo{person}{A. Poelstra},
  \bibinfo{person}{P. Wuille}, {and} \bibinfo{person}{G. Maxwell}.}
  \bibinfo{year}{2018}\natexlab{}.
\newblock \showarticletitle{Bulletproofs: Short proofs for confidential
  transactions and more}. In \bibinfo{booktitle}{\emph{IEEE symposium on
  security and privacy (SP)}}.
\newblock


\bibitem[\protect\citeauthoryear{Chau and Elbassioni}{Chau and
  Elbassioni}{2018}]%
        {CE17sharing}
\bibfield{author}{\bibinfo{person}{Chi-Kin Chau} {and} \bibinfo{person}{Khaled
  Elbassioni}.} \bibinfo{year}{2018}\natexlab{}.
\newblock \showarticletitle{Quantifying Inefficiency of Fair Cost-Sharing
  Mechanisms for Sharing Economy}.
\newblock \bibinfo{journal}{\emph{IEEE Trans. Control of Network System}}
  \bibinfo{volume}{5} (\bibinfo{date}{Dec} \bibinfo{year}{2018}),
  \bibinfo{pages}{1809--1818}.
\newblock
Issue 4.


\bibitem[\protect\citeauthoryear{Chau, Xu, Bow, and Elbassioni}{Chau
  et~al\mbox{.}}{2019}]%
        {p2p19}
\bibfield{author}{\bibinfo{person}{Sid Chi-Kin Chau}, \bibinfo{person}{Jiajia
  Xu}, \bibinfo{person}{Wilson Bow}, {and} \bibinfo{person}{Khaled
  Elbassioni}.} \bibinfo{year}{2019}\natexlab{}.
\newblock \showarticletitle{Peer-to-Peer Energy Sharing: Effective Cost-Sharing
  Mechanisms and Social Efficiency}. In \bibinfo{booktitle}{\emph{ACM Intl.
  Conf. on Future Energy Systems (e-Energy)}}.
\newblock


\bibitem[\protect\citeauthoryear{Chen, Pastro, and Raykova}{Chen
  et~al\mbox{.}}{2018}]%
        {cpr18spdz}
\bibfield{author}{\bibinfo{person}{Valerie Chen}, \bibinfo{person}{Valerio
  Pastro}, {and} \bibinfo{person}{Mariana Raykova}.}
  \bibinfo{year}{2018}\natexlab{}.
\newblock \showarticletitle{Secure Computation for Machine Learning With SPDZ}.
  In \bibinfo{booktitle}{\emph{Annual Conference on Neural Information
  Processing Systems (NeurIPS)}}.
\newblock


\bibitem[\protect\citeauthoryear{CoinDesk}{CoinDesk}{2021}]%
        {coindesk}
\bibfield{author}{\bibinfo{person}{CoinDesk}.} \bibinfo{year}{2021}\natexlab{}.
\newblock
  \bibinfo{howpublished}{\url{https://www.coindesk.com/price/ethereum}}.
  (\bibinfo{year}{2021}).
\newblock


\bibitem[\protect\citeauthoryear{Cramer, Damg{\aa}rd, and Nielsen}{Cramer
  et~al\mbox{.}}{2001}]%
        {cramer2001multiparty}
\bibfield{author}{\bibinfo{person}{Ronald Cramer}, \bibinfo{person}{Ivan
  Damg{\aa}rd}, {and} \bibinfo{person}{Jesper~B Nielsen}.}
  \bibinfo{year}{2001}\natexlab{}.
\newblock \showarticletitle{Multiparty computation from threshold homomorphic
  encryption}. In \bibinfo{booktitle}{\emph{Intl. conference on the theory and
  applications of cryptographic techniques}}.
\newblock


\bibitem[\protect\citeauthoryear{Cramer, Damg{\aa}rd, and Nielsen}{Cramer
  et~al\mbox{.}}{2015}]%
        {cramer2015secure}
\bibfield{author}{\bibinfo{person}{Ronald Cramer}, \bibinfo{person}{Ivan~Bjerre
  Damg{\aa}rd}, {and} \bibinfo{person}{Jesper~Buus Nielsen}.}
  \bibinfo{year}{2015}\natexlab{}.
\newblock \bibinfo{booktitle}{\emph{Secure Multiparty Computation and Secret
  Sharing}}.
\newblock \bibinfo{publisher}{Cambridge University Press}.
\newblock
\newblock
\shownote{Cambridge Books Online.}


\bibitem[\protect\citeauthoryear{Damg{\aa}rd, Keller, Larraia, Pastro, Scholl,
  and Smart}{Damg{\aa}rd et~al\mbox{.}}{2013}]%
        {dklpss13spdz}
\bibfield{author}{\bibinfo{person}{Ivan Damg{\aa}rd}, \bibinfo{person}{Marcel
  Keller}, \bibinfo{person}{Enrique Larraia}, \bibinfo{person}{Valerio Pastro},
  \bibinfo{person}{Peter Scholl}, {and} \bibinfo{person}{Nigel~P. Smart}.}
  \bibinfo{year}{2013}\natexlab{}.
\newblock \showarticletitle{Practical Covertly Secure MPC for Dishonest
  Majority - or: Breaking the SPDZ Limits}. In
  \bibinfo{booktitle}{\emph{European Symposium on Research in Computer Security
  (ESORICS)}}.
\newblock


\bibitem[\protect\citeauthoryear{Du and Atallah}{Du and Atallah}{2001}]%
        {du2001secure}
\bibfield{author}{\bibinfo{person}{Wenliang Du} {and}
  \bibinfo{person}{Mikhail~J Atallah}.} \bibinfo{year}{2001}\natexlab{}.
\newblock \showarticletitle{Secure multi-party computation problems and their
  applications: a review and open problems}. In \bibinfo{booktitle}{\emph{the
  Workshop on New Security Paradigms}}.
\newblock


\bibitem[\protect\citeauthoryear{Dwork, McSherry, Nissim, and Smith}{Dwork
  et~al\mbox{.}}{2006}]%
        {dwork2006calibrating}
\bibfield{author}{\bibinfo{person}{Cynthia Dwork}, \bibinfo{person}{Frank
  McSherry}, \bibinfo{person}{Kobbi Nissim}, {and} \bibinfo{person}{Adam
  Smith}.} \bibinfo{year}{2006}\natexlab{}.
\newblock \showarticletitle{Calibrating noise to sensitivity in private data
  analysis}. In \bibinfo{booktitle}{\emph{Theory of cryptography conference}}.
  Springer.
\newblock


\bibitem[\protect\citeauthoryear{Edalat, Motani, Walrand, and Huang}{Edalat
  et~al\mbox{.}}{2014}]%
        {control14}
\bibfield{author}{\bibinfo{person}{Neda Edalat}, \bibinfo{person}{Mehul
  Motani}, \bibinfo{person}{Jean Walrand}, {and} \bibinfo{person}{Longbo
  Huang}.} \bibinfo{year}{2014}\natexlab{}.
\newblock \showarticletitle{Control of systems that store renewable energy}. In
  \bibinfo{booktitle}{\emph{ACM Intl. Conf. on Future Energy Systems
  (e-Energy)}}.
\newblock


\bibitem[\protect\citeauthoryear{Gas and Company}{Gas and Company}{2018}]%
        {vnmPGnE}
\bibfield{author}{\bibinfo{person}{Pacific Gas} {and} \bibinfo{person}{Electric
  Company}.} \bibinfo{year}{2018}\natexlab{}.
\newblock \showarticletitle{Understanding the Virtual Net Energy Metering
  Program A guide for statements and bills}.
\newblock  (\bibinfo{year}{2018}).
\newblock


\bibitem[\protect\citeauthoryear{Gelman and Bird}{Gelman and Bird}{2014}]%
        {nrel14}
\bibfield{author}{\bibinfo{person}{J.~Heeter~R. Gelman} {and}
  \bibinfo{person}{L. Bird}.} \bibinfo{year}{2014}\natexlab{}.
\newblock \bibinfo{booktitle}{\emph{Status of Net Metering: Assessing the
  Potential to Reach Program Caps}}.
\newblock \bibinfo{type}{{T}echnical {R}eport}.
\newblock


\bibitem[\protect\citeauthoryear{Goldreich}{Goldreich}{1998}]%
        {goldreich1998secure}
\bibfield{author}{\bibinfo{person}{Oded Goldreich}.}
  \bibinfo{year}{1998}\natexlab{}.
\newblock \showarticletitle{Secure multi-party computation}.
\newblock \bibinfo{journal}{\emph{Manuscript. Preliminary version}}
  \bibinfo{volume}{78} (\bibinfo{year}{1998}).
\newblock


\bibitem[\protect\citeauthoryear{Gorenflo, Golab, and Keshav}{Gorenflo
  et~al\mbox{.}}{2019}]%
        {ggk19block}
\bibfield{author}{\bibinfo{person}{Christian Gorenflo}, \bibinfo{person}{Lukasz
  Golab}, {and} \bibinfo{person}{Srinivasan Keshav}.}
  \bibinfo{year}{2019}\natexlab{}.
\newblock \showarticletitle{Using a Blockchain to Mitigate Trust in Electric
  Vehicle Charging}. In \bibinfo{booktitle}{\emph{ACM Intl. Conf. on Future
  Energy Systems (e-Energy)}}.
\newblock


\bibitem[\protect\citeauthoryear{Hajiesmaili, Chen, Mallada, and
  Chau}{Hajiesmaili et~al\mbox{.}}{2017}]%
        {crowd17}
\bibfield{author}{\bibinfo{person}{Mohammad~H. Hajiesmaili},
  \bibinfo{person}{Minghua Chen}, \bibinfo{person}{Enrique Mallada}, {and}
  \bibinfo{person}{Chi-Kin Chau}.} \bibinfo{year}{2017}\natexlab{}.
\newblock \showarticletitle{Crowd-Sourced Storage-Assisted Demand Response in
  Microgrids}. In \bibinfo{booktitle}{\emph{ACM Intl. Conf. on Future Energy
  Systems (e-Energy)}}.
\newblock


\bibitem[\protect\citeauthoryear{Hazay and Lindell}{Hazay and Lindell}{2010}]%
        {HazayL10}
\bibfield{author}{\bibinfo{person}{Carmit Hazay} {and} \bibinfo{person}{Yehuda
  Lindell}.} \bibinfo{year}{2010}\natexlab{}.
\newblock \bibinfo{booktitle}{\emph{Efficient Secure Two-Party Protocols -
  Techniques and Constructions}}.
\newblock \bibinfo{publisher}{Springer}.
\newblock


\bibitem[\protect\citeauthoryear{Huang, Zhu, Gu, and Li}{Huang
  et~al\mbox{.}}{2016}]%
        {shepherd16}
\bibfield{author}{\bibinfo{person}{Zhichuan Huang}, \bibinfo{person}{Ting Zhu},
  \bibinfo{person}{Yu Gu}, {and} \bibinfo{person}{Yanhua Li}.}
  \bibinfo{year}{2016}\natexlab{}.
\newblock \showarticletitle{Shepherd: sharing energy for privacy preserving in
  hybrid AC-DC microgrids}. In \bibinfo{booktitle}{\emph{ACM Intl. Conf. on
  Future Energy Systems (e-Energy)}}.
\newblock


\bibitem[\protect\citeauthoryear{Knirsch, Brunner, Unterweger, and
  Engel}{Knirsch et~al\mbox{.}}{2020}]%
        {kbue20gecko}
\bibfield{author}{\bibinfo{person}{Fabian Knirsch}, \bibinfo{person}{Clemens
  Brunner}, \bibinfo{person}{Andreas Unterweger}, {and}
  \bibinfo{person}{Dominik Engel}.} \bibinfo{year}{2020}\natexlab{}.
\newblock \showarticletitle{Decentralized and permission-less green energy
  certificates with GECKO}.
\newblock \bibinfo{journal}{\emph{Energy Informatics}} \bibinfo{volume}{3},
  \bibinfo{number}{2} (\bibinfo{year}{2020}).
\newblock


\bibitem[\protect\citeauthoryear{Laforet, Buchmann, and Bohm}{Laforet
  et~al\mbox{.}}{2016}]%
        {privbat16}
\bibfield{author}{\bibinfo{person}{Fabian Laforet}, \bibinfo{person}{Erik
  Buchmann}, {and} \bibinfo{person}{Klemens Bohm}.}
  \bibinfo{year}{2016}\natexlab{}.
\newblock \showarticletitle{Towards provable privacy guarantees using
  rechargeable energy-storage devices}. In \bibinfo{booktitle}{\emph{ACM Intl.
  Conf. on Future Energy Systems (e-Energy)}}.
\newblock


\bibitem[\protect\citeauthoryear{Language}{Language}{2020}]%
        {solidity}
\bibfield{author}{\bibinfo{person}{Solidity~Programming Language}.}
  \bibinfo{year}{2020}\natexlab{}.
\newblock \bibinfo{howpublished}{\url{https://docs.soliditylang.org}}.
  (\bibinfo{year}{2020}).
\newblock


\bibitem[\protect\citeauthoryear{Lee, Shenoy, Ramamritham, and Irwin}{Lee
  et~al\mbox{.}}{2018}]%
        {vsolar18}
\bibfield{author}{\bibinfo{person}{Stephen Lee}, \bibinfo{person}{Prashant
  Shenoy}, \bibinfo{person}{Krithi Ramamritham}, {and} \bibinfo{person}{David
  Irwin}.} \bibinfo{year}{2018}\natexlab{}.
\newblock \showarticletitle{vSolar: Virtualizing Community Solar and Storage
  for Energy Sharing}. In \bibinfo{booktitle}{\emph{ACM Intl. Conf. on Future
  Energy Systems (e-Energy)}}.
\newblock


\bibitem[\protect\citeauthoryear{Liu, Zhang, Kang, Kirschen, and Xia}{Liu
  et~al\mbox{.}}{2017}]%
        {cloud17}
\bibfield{author}{\bibinfo{person}{Jingkun Liu}, \bibinfo{person}{Ning Zhang},
  \bibinfo{person}{Chongqing Kang}, \bibinfo{person}{Daniel Kirschen}, {and}
  \bibinfo{person}{Qing Xia}.} \bibinfo{year}{2017}\natexlab{}.
\newblock \showarticletitle{Cloud energy storage for residential and small
  commercial consumers: A business case study}.
\newblock \bibinfo{journal}{\emph{Applied Energy}}  \bibinfo{volume}{188}
  (\bibinfo{year}{2017}), \bibinfo{pages}{226--236}.
\newblock


\bibitem[\protect\citeauthoryear{Lyu, Chau, Wang, and Zheng}{Lyu
  et~al\mbox{.}}{2020}]%
        {lcwz20privsharing}
\bibfield{author}{\bibinfo{person}{Lingjuan Lyu}, \bibinfo{person}{Sid Chi-Kin
  Chau}, \bibinfo{person}{Nan Wang}, {and} \bibinfo{person}{Yifeng Zheng}.}
  \bibinfo{year}{2020}\natexlab{}.
\newblock \showarticletitle{Cloud-based Privacy-Preserving Collaborative
  Consumption for Sharing Economy}.
\newblock \bibinfo{journal}{\emph{IEEE Trans. Cloud Computing}}
  (\bibinfo{year}{2020}).
\newblock


\bibitem[\protect\citeauthoryear{Mengelkamp, Garttner, Rock, Kessler, Orsini,
  and Weinhardt}{Mengelkamp et~al\mbox{.}}{2018}]%
        {m18brooklyn}
\bibfield{author}{\bibinfo{person}{Esther Mengelkamp},
  \bibinfo{person}{Johannes Garttner}, \bibinfo{person}{Kerstin Rock},
  \bibinfo{person}{Scott Kessler}, \bibinfo{person}{Lawrence Orsini}, {and}
  \bibinfo{person}{Christof Weinhardt}.} \bibinfo{year}{2018}\natexlab{}.
\newblock \showarticletitle{Designing microgrid energy markets: A case study:
  The Brooklyn Microgrid}.
\newblock \bibinfo{journal}{\emph{Applied Energy}}  \bibinfo{volume}{210}
  (\bibinfo{year}{2018}), \bibinfo{pages}{870--880}.
\newblock


\bibitem[\protect\citeauthoryear{Mishra, Irwin, Shenoy, Kurose, and Zhu}{Mishra
  et~al\mbox{.}}{2012}]%
        {smartCharge12}
\bibfield{author}{\bibinfo{person}{Aditya Mishra}, \bibinfo{person}{David
  Irwin}, \bibinfo{person}{Prashant Shenoy}, \bibinfo{person}{Jim Kurose},
  {and} \bibinfo{person}{Ting Zhu}.} \bibinfo{year}{2012}\natexlab{}.
\newblock \showarticletitle{SmartCharge: cutting the electricity bill in smart
  homes with energy storage}. In \bibinfo{booktitle}{\emph{ACM Intl. Conf. on
  Future Energy Systems (e-Energy)}}.
\newblock


\bibitem[\protect\citeauthoryear{Monero}{Monero}{2021}]%
        {monero}
\bibfield{author}{\bibinfo{person}{Monero}.} \bibinfo{year}{2021}\natexlab{}.
\newblock \bibinfo{howpublished}{\url{http://getmonero.org}}.
  (\bibinfo{year}{2021}).
\newblock


\bibitem[\protect\citeauthoryear{Paper}{Paper}{2014}]%
        {ethereum}
\bibfield{author}{\bibinfo{person}{The Ethereum~Yellow Paper}.}
  \bibinfo{year}{2014}\natexlab{}.
\newblock
  \bibinfo{howpublished}{\url{https://ethereum.github.io/yellowpaper/paper.pdf}}.
    (\bibinfo{year}{2014}).
\newblock


\bibitem[\protect\citeauthoryear{Quantiki}{Quantiki}{2020}]%
        {cointossing}
\bibfield{author}{\bibinfo{person}{Quantiki}.} \bibinfo{year}{2020}\natexlab{}.
\newblock \bibinfo{howpublished}{\url{https://quantiki.org/wiki/coin-tossing}}.
    (\bibinfo{year}{2020}).
\newblock


\bibitem[\protect\citeauthoryear{Shaw-Williams and Susilawati}{Shaw-Williams
  and Susilawati}{2020}]%
        {vnm20}
\bibfield{author}{\bibinfo{person}{Damian Shaw-Williams} {and}
  \bibinfo{person}{Connie Susilawati}.} \bibinfo{year}{2020}\natexlab{}.
\newblock \showarticletitle{A techno-economic evaluation of Virtual Net
  Metering for the Australian community housing sector}.
\newblock \bibinfo{journal}{\emph{Applied Energy}}  \bibinfo{volume}{261}
  (\bibinfo{year}{2020}).
\newblock


\bibitem[\protect\citeauthoryear{Shi, Chan, Rieffel, Chow, and Song}{Shi
  et~al\mbox{.}}{2011}]%
        {shi2011privacy}
\bibfield{author}{\bibinfo{person}{Elaine Shi}, \bibinfo{person}{HTH Chan},
  \bibinfo{person}{Eleanor Rieffel}, \bibinfo{person}{Richard Chow}, {and}
  \bibinfo{person}{Dawn Song}.} \bibinfo{year}{2011}\natexlab{}.
\newblock \showarticletitle{Privacy-preserving aggregation of time-series
  data}. In \bibinfo{booktitle}{\emph{Annual Network \& Distributed System
  Security Symposium (NDSS)}}.
\newblock


\bibitem[\protect\citeauthoryear{Station}{Station}{2020}]%
        {ethgasstation}
\bibfield{author}{\bibinfo{person}{ETH~Gas Station}.}
  \bibinfo{year}{2020}\natexlab{}.
\newblock \bibinfo{howpublished}{\url{https://ethgasstation.info}}.
  (\bibinfo{year}{2020}).
\newblock


\bibitem[\protect\citeauthoryear{Suite}{Suite}{2020}]%
        {truffle}
\bibfield{author}{\bibinfo{person}{Truffle Suite}.}
  \bibinfo{year}{2020}\natexlab{}.
\newblock \bibinfo{howpublished}{\url{https://www.trufflesuite.com}}.
  (\bibinfo{year}{2020}).
\newblock


\bibitem[\protect\citeauthoryear{Wang, Chau, and Zhou}{Wang
  et~al\mbox{.}}{2021}]%
        {chau21blockchain}
\bibfield{author}{\bibinfo{person}{Nan Wang}, \bibinfo{person}{Sid Chi-Kin
  Chau}, {and} \bibinfo{person}{Yue Zhou}.} \bibinfo{year}{2021}\natexlab{}.
\newblock \showarticletitle{Privacy-Preserving Energy Storage Sharing with
  Blockchain}. In \bibinfo{booktitle}{\emph{Proc. of ACM e-Energy}}.
\newblock


\end{thebibliography}

\appendix

\section*{Appendix}

\section{Proofs} \label{sec:append1}

\begin{customthm}{1}
If $\big({x}^-(t), {y}(t)\big)_{t=1}^T$ is an optimal solution of ${\tt (P2)}$, then $\big(({x}^-_i(t), {y}_i(t))_{i=1}^N\big)_{t=1}^T$, where ${x}^-_i(t) =  \frac{a_i(t)}{a(t)} \cdot {x}^-(t)$ and ${y}_i(t) = \frac{a_i(t)}{a(t)} \cdot {y}(t)$, is an optimal solution of ${\tt (P1)}$.
\end{customthm}
\begin{proof}
Since $\big({x}^-(t), {y}(t) \big)_{t=1}^T$ is an optimal solution of ${\tt (P2)}$, it satisfies the condition ${x}^-(t) +  {y}(t)   = a(t)$. 

We let ${x}^-_i(t) =  \frac{a_i(t)}{a(t)} \cdot {x}^-(t)$ and ${y}_i(t) = \frac{a_i(t)}{a(t)} \cdot {y}(t)$, then $\big({x}^-_i(t), {y}_i(t) \big)_{t=1}^T$ satisfies ${x}^-_i(t) +  {y}_i(t)   = a_i(t)$ for all $i \in \{1, ..., N\}$ and $t \in \{1, ..., T\}$, and hence, is a feasible solution of ${\tt (P1)}$. 

Next, we argue that $\big(({x}^-_i(t), {y}_i(t))_{i=1}^N\big)_{t=1}^T$ is an optimal solution of ${\tt (P1)}$ by contradiction. Suppose that there exists a better solution $\big(({x'}^-_i(t), {y'}_i(t))_{i=1}^N\big)_{t=1}^T$ with a lower total cost in ${\tt (P2)}$. Then $\big({x}^-(t), {y}(t)\big)_{t=1}^T$ is not an optimal solution, because we can find another better solution by considering $\big(\sum_{i=1}^N{x'}^-_i(t), \sum_{i=1}^N {y'}_i(t) \big)_{t=1}^T$ instead, which is also a feasible solution of ${\tt (P2)}$. This will violate the optimality of $\big({x}^-(t), {y}(t)\big)$. 
\end{proof}

\medskip

\begin{customthm}{2}
If $\big({x}^+(t), {x}^-(t)\big)_{t=1}^T$ is an optimal solution of ${\tt (P2)}$ and let ${x}^-_i(t) =  \frac{a_i(t)}{a(t)} \cdot {x}^-(t)$ and ${y}_i(t) = \frac{a_i(t)}{a(t)} \cdot {y}(t)$, then proportional and egalitarian cost-sharing schemes are individually rational.

Let $\hat{\tt p}(t) \triangleq \frac{{x}^-(t) \cdot p(t)}{a(t)}$ and ${\tt Cost}_{\tt org} \triangleq \sum_{t=1}^T {x}^-(t) \cdot p(t)$. The proportional and egalitarian cost-sharing payments are given as follows:
\begin{equation}
\begin{cases}
P_i^{\tt pp} =  \frac{{\tt Cost}_{\tt ess}}{{\tt Cost}_{\tt org}} \cdot \sum_{t=1}^T a_i(t) \cdot \hat{\tt p}(t), \\  
P_i^{\tt ega} =  \sum_{t=1}^T a_i(t) \cdot \hat{\tt p}(t)  - \frac{{\tt Cost}_{\tt org}  - {\tt Cost}_{\tt ess}}{N} 
\end{cases}
\end{equation}
\end{customthm}
\begin{proof}
For proportional cost sharing, $U_i$'s saving will be 
\begin{align}
 \Delta_i^{\tt pp}  = & \sum_{t=1}^T {x}^-_i(t) \cdot p(t) - \sum_{t=1}^T \big( p(t) + {\tt p}_{\tt s} \big) \cdot {x}^+(t) \cdot \frac{\sum_{t=1}^T {x}^-_i(t) \cdot p(t)}{\sum_{t=1}^T {x}^-(t) \cdot p(t)} \\
 = & \Big(  \sum_{t=1}^T {x}^-(t) \cdot p(t) - \sum_{t=1}^T \big( p(t) + {\tt p}_{\tt s} \big) \cdot {x}^+(t)  \Big) \cdot \frac{\sum_{t=1}^T {x}^-_i(t) \cdot p(t)}{\sum_{t=1}^T {x}^-(t) \cdot p(t)}
\end{align}

Next, we show $\sum_{t=1}^T {x}^-(t) \cdot p(t) \ge \sum_{t=1}^T \big( p(t) + {\tt p}_{\tt s} \big) \cdot {x}^+(t) $ by contradiction. Suppose $\sum_{t=1}^T {x}^-(t) \cdot p(t) < \sum_{t=1}^T \big( p(t) + {\tt p}_{\tt s} \big) \cdot {x}^+(t)$, then $\big({x}^+(t), {x}^-(t)\big)_{t=1}^T$ is not an optimal solution of ${\tt (P2)}$ because one can always find a better solution by not charging energy storage according to ${x}^+(t)$. Instead, drawing energy at the time it is needed will only cost $\sum_{t=1}^T {x}^-(t) \cdot p(t)$, which is cheaper than the cost of charging and subsequently discharging from energy storage ($\sum_{t=1}^T \big( p(t) + {\tt p}_{\tt s} \big) \cdot {x}^+(t)$). Hence, we conclude that $\sum_{t=1}^T {x}^-(t) \cdot p(t) \ge \sum_{t=1}^T \big( p(t) + {\tt p}_{\tt s} \big) \cdot {x}^+(t) $ and $ \Delta_i^{\tt pp}  \ge 0$.

For egalitarian cost sharing, $U_i$'s saving will be 
\begin{equation}
 \Delta_i^{\tt ega} = \frac{\sum_{t=1}^T {x}^-(t) \cdot p(t) - \sum_{t=1}^T \big( p(t) + {\tt p}_{\tt s} \big) \cdot {x}^+(t)}{N}
\end{equation}

Following a similar approach by contradiction, we can similarly show that $\Delta_i^{\tt ega}  \ge 0$.
\end{proof}

\medskip

\begin{customthm}{3} 
Consider a multi-transaction ${\tt mtx} = ({\tt ad}_i, {\tt ad}_{\tt ess},$ ${\tt val}_i)_{i=1}^N$, where ${\tt val}_i$ may be negative. Namely, every ${\tt ad}_i$ pays to the energy storage operator ${\tt ad}_{\tt ess}$. If $\sum_{i=1}^N {\tt val}_i > 0$, ${\tt mtx}$ can be handled on a blockchain by the following transaction operations:
\begin{align}
{\tt Bal}({\tt ad}_{i}) \leftarrow & {\tt Bal}({\tt ad}_{i}) - {\tt val}_i, \mbox{for all\ }i \label{eqn:mtx-ud}\\
{\tt Bal}({\tt ad}_{\tt ess}) \leftarrow & {\tt Bal}({\tt ad}_{\tt ess}) + \sum_{i=1}^N 	{\tt val}_i
\end{align}
\end{customthm}
\begin{proof}
It is straightforward to see that Eqn.~\raf{eqn:mtx-ud} applies to the case when ${\tt val}_i$ is negative. As long as $\sum_{i=1}^N {\tt val}_i > 0$, there is no net out-going payment from ${\tt ad}_{\tt ess}$. Therefore, ${\tt mtx}$ can be handled properly.
\end{proof}

\section{Zero-knowledge Proofs of Knowledge} \label{sec:append2}

\subsection{Zero-knowledge Proof of Commitment ({\tt zkpCm})}

Given ${\tt Cm}(x, {\tt r})$, a prover wants to convince a verifier of the knowledge of $(x, {\tt r})$. We can apply $\Sigma$-protocol as follows:

\begin{enumerate}

\item The prover randomly generates $(x', {\tt r}') \in {\mathbb Z}^2_p$ and sends the commitment ${\tt Cm}(x', {\tt r}')$ to the verifier.

\item The verifier sends a random challenge $\beta \in {\mathbb Z}_p$ to the prover. 

\item The prover replies with $z_x = x' + \beta \cdot x$ and $z_{\tt r} = {\tt r}' + \beta \cdot {\tt r}$.

\item The verifier checks whether $g^{z_x} \cdot h^{z_{\tt r}} \overset{?}{=}  {\tt Cm}(x', {\tt r}') \cdot {\tt Cm}(x, {\tt r})^\beta$.

\end{enumerate}
Denote a zero-knowledge proof of commitment for ${\tt Cm}(x, {\tt r})$ by ${\tt zkpCm}[x]$.
 
\subsection{Zero-knowledge Proof of Summation}

Given commitments $\big( {\tt Cm}(x_1, {\tt r}_1),..., {\tt Cm}(x_n, {\tt r}_n) \big)$ and $y$, a prover wants to convince a verifier of the knowledge of $y = \sum_{i=1}^n x_i$ without revealing $(x_1, ..., x_n)$. We can apply $\Sigma$-protocol as follows:

\begin{enumerate}

\item The prover randomly generates ${\tt r}'\in {\mathbb Z}_p$ and sends the commitment ${\tt Cm}(0, {\tt r}')$ to the verifier.

\item The verifier sends a random challenge $\beta\in {\mathbb Z}_p$ to the prover. 

\item The prover replies with $z_{\tt r} = {\tt r}' + \beta \cdot \sum_{i=1}^n {\tt r}_i$.

\item The verifier checks whether $g^{\beta y}\cdot h^{z_{\tt r}} \overset{?}{=}  {\tt Cm}(0, {\tt r}')\cdot\prod_{i=1}^n {\tt Cm}(x_i, {\tt r}_i)^\beta$

\end{enumerate}
Denote a zero-knowledge proof of summation for $\big( {\tt Cm}(x_i, {\tt r}_i)\big)_{i=1}^n$ by ${\tt zkpSum}[y, (x_i)_{i=1}^n]$.

\subsection{Zero-knowledge Proof of Membership}

Given a set ${\mathcal X} = \{x_1, ..., x_n \}$ and ${\tt Cm}(x, {\tt r})$, a prover wants to convince a verifier of the knowledge of $x \in {\mathcal X}$ without revealing $x$. We can apply $\Sigma$-protocol as follows:

\begin{enumerate}

\item Suppose $x=x_i \in {\mathcal X}$. The prover first randomly generates $(x'_j, {\tt r}'_j) \in {\mathbb Z}_p$ and computes the commitment ${\tt Cm}(x'_j, {\tt r}'_j)$ for all $j\in\{1,...,n\}$. Then, the prover randomly generates $\beta_j\in {\mathbb Z}_p$ for each $j\in\{1,...,n\}\backslash\{i\}$, and computes
\[
z_{x_j} = 
\begin{cases}
x'_j + (x_i - x_j) \beta_j,& \mbox{if\ } j\in\{1,...,n\}\backslash\{i\}\\
x'_i, & \mbox{if\ } j = i\\
\end{cases}
\] 
Next, the prover sends $({\tt Cm}(x'_j, {\tt r}'_j), z_{x_j})_{j=1}^n$ to the verifier.

\item The verifier sends a random challenge $\beta\in {\mathbb Z}_p$ to the prover. 

\item The prover sets $\beta_i = \beta - \sum_{j\ne i}\beta_j$, then computes $z_{{\tt r}_j} = {\tt r}'_j + {\tt r} \cdot \beta_j$ for all $j\in\{1,...,n\}$, and sends $(\beta_j, z_{{\tt r}_j})_{j=1}^n$ to the verifier.

\item The verifier checks whether $\beta \overset{?}{=} \sum_{i=1}^n \beta_j$ and 
\[
g^{z_{x_j}} \cdot h^{z_{{\tt r}_j}} \overset{?}{=} {\tt Cm}(x'_j, {\tt r}'_j) \cdot \Big( \frac{{\tt Cm}(x, {\tt r})}{g^{x_j}} \Big)^{\beta_j} \mbox{for all\ } j\in\{1,...,n\}
\]

\end{enumerate}
Denote a zero-knowledge proof of membership for $x \in {\mathcal X}$ by ${\tt zkpMbs}[x, {\mathcal X}]$.

\subsection{Zero-knowledge Proof of Non-Negativity}

Given ${\tt Cm}(x, {\tt r})$, a prover wants to convince a verifier of the knowledge of $x \ge 0$ without revealing $x$. Suppose $x < 2^m$. We aim to prove there exist $(b_1, ..., b_m)$ such that $b_i \in \{0, 1\}$ for $i \in \{0, ..., m\}$ and $\sum_{i=1}^{m}b_i \cdot 2^{i-1} = x$. We can apply $\Sigma$-protocol as follows:
\begin{enumerate}

\item The prover sends $({\tt Cm}(b_i, {\tt r}_i))_{i=1}^{m}$ to the verifier, and provides ${\tt zkpMbs}[b_i, \{0,1\}]$ for each $b_i$ to prove that $b_i \in \{0, 1\}$. Also, the prover randomly generates ${\tt r}'\in {\mathbb Z}_p$ and sends the commitment ${\tt Cm}(0, {\tt r}')$ to the verifier.

\item The verifier sends a random challenge $\beta\in {\mathbb Z}_p$ to the prover. 

\item The prover replies with $z_{\tt r} = {\tt r}' + \beta \cdot (\sum_{i=1}^m {\tt r}_i \cdot 2^{i-1} - {\tt r})$.

\item The verifier checks whether $h^{z_{\tt r}} \overset{?}{=}  {\tt Cm}(0,{\tt r}') \cdot {\tt Cm}(x, {\tt r})^{-\beta}\cdot\prod_{i=1}^m {\tt Cm}(b_i, {\tt r}_i)^{\beta\cdot 2^{i-1}}$.

\end{enumerate}
Denote a zero-knowledge proof of $x \geq 0$ by ${\tt zkpNN}[x]$.

\subsection{Security Proof} \label{sec:zkpproofs}

It is straightforward to prove the completeness of these protocols. We will provide detailed proofs on their soundness and honest-verifier zero-knowledge properties below.

\subsubsection{Soundness Proof} 

Proving soundness is equivalent to showing that there exists a {\em knowledge extractor} who makes the prover successfully answer two random given challenges $\beta_1$ and $\beta_2$:
\begin{itemize}
\item{({\tt zkpCm})} For $x$, we have $(z_x' = x' + \beta_1 \cdot x$, $z_x'' = x' + \beta_2 \cdot x)$. For $r$, we have $(z_r' = r' + \beta_1 \cdot r$, $z_r'' = r' + \beta_2 \cdot r)$. Then we obtain $x=\dfrac{z_x'-z_x''}{\beta_1 - \beta_2}$ and $r=\dfrac{z_r'-z_r''}{\beta_1 - \beta_2}$. Finally, we can check that $g^x h^r={\tt Cm}(x, r)$.

\item{({\tt zkpSum})} Let $z_1 = r' + \beta_1 \cdot \sum_{i=1}^N r_i$ and $z_2 = r' + \beta_2 \cdot \sum_{i=1}^N r_i$. Then we have $\sum_{i=1}^N r_i=\dfrac{z_1-z_2}{\beta_1 - \beta_2}$. Finally, we can check that $g^y h^{\sum_{i=1}^N r_i}=\prod_{i=1}^n {\tt Cm}(x_i, {\tt r}_i)$ to prove that $y=\sum_{i=1}^N x_i$. 

\item{({\tt zkpMbs})} Let $z'_{r_j} = r'_j + \beta_1 \cdot r$ and $z''_{r_j} = r' + \beta_2 \cdot r$. Then we have $r=\dfrac{z'_{r_j}-z''_{r_j}}{\beta_1 - \beta_2}$. Finally, we can check for each $j$ that $g^{z_{x_j}} h^r=\dfrac{{\tt Cm}(x, {\tt r})}{g^{x_j}}$ to prove that $x \in {\mathcal X}$. 

\item{({\tt zkpNN})} Let $z'_r = r' + \beta_1 \cdot (\sum_{i=1}^m {\tt r}_i \cdot 2^{i-1} - {\tt r})$ and $z''_r = r' + \beta_2 \cdot (\sum_{i=1}^m {\tt r}_i \cdot 2^{i-1} - {\tt r})$. Then we have $(\sum_{i=1}^m {\tt r}_i \cdot 2^{i-1} - {\tt r})=\dfrac{z'_{r_j}-z''_{r_j}}{\beta_1 - \beta_2}$. Finally, we can check that $h^{\sum_{i=1}^m {\tt r}_i \cdot 2^{i-1} - {\tt r}}=\prod_{i=1}^m {\tt Cm}(b_i, {\tt r}_i)^{2^{i-1}} \cdot {\tt Cm}(x, {\tt r})^{-1}$ to prove that $x=\sum_{i=1}^N b_i \cdot 2^{i-1} \geq 0$. 

\end{itemize}

\subsubsection{Honest-verifier Zero-knowledge Proof} 

It suffices to show that there exists a {\em simulator} who can produce another set of zero-knowledge proofs that are computationally indistinguishable from a given set of zero-knowledge proofs:
\begin{itemize}
\item{({\tt zkpCm})} ${\tt Cm}(x', r')=g^{z_x} \cdot h^{z_r} \cdot {\tt Cm}(x, r)^{-\beta}$.

\item{({\tt zkpSum})} ${\tt Cm}(0, {\tt r}') =g^{\beta y} \cdot h^{z_{\tt r}} \cdot \prod_{i=1}^N {\tt Cm}(x_i, {\tt r}_i)^{-\beta}$.

\item{({\tt zkpMbs})} ${\tt Cm}(x'_j, {\tt r}'_j) = g^{z_{x_j}} \cdot h^{z_{{\tt r}_j}} \cdot \Big( \frac{{\tt Cm}(x, {\tt r})}{g^{x_j}} \Big)^{-\beta_j} \mbox{for all\ } j\in\{1,...,n\}$.

\item{({\tt zkpNN})} ${\tt Cm}(0,{\tt r}') = h^{z_{\tt r}} \cdot {\tt Cm}(x, {\tt r})^{\beta}\cdot\prod_{i=1}^m {\tt Cm}(b_i, {\tt r}_i)^{-\beta\cdot 2^{i-1}}$.

\end{itemize}
\section{SPDZ Protocol} \label{sec:append3}

In the following, we present a simplified version of SPDZ for the clarity of exposition. The full version can be found in \cite{cramer2015secure, dklpss13spdz}. 

There are three phases in SPDZ protocol: (1) pre-processing phase, (2) online phase, and (3) output and validation phase. We write $\langle x \rangle$ as a {\em secretly shared} number, meaning that there is a vector $(x_1, ..., x_n)$, such that each party $i$ knows only $x_i$. To reveal secretly shared number $\langle x \rangle$, each party $i$ broadcasts $x_i$ to other parties. Then each party can reconstruct $x = \sum_{i=1}^n x_i$. We write $\llangle  x \rrangle$ meaning that both $\langle  x \rangle$ and the respective MAC $\langle \gamma(x) \rangle$ are secretly shared.

\subsection{Online Phase}

In the online phase, the parties can jointly compute an arithmetic circuit, consisting of additions and multiplications with secretly shared input numbers.

\subsubsection{Addition}
\

Given secretly shared $\langle x \rangle$ and $\langle y \rangle$, and a public known constant $c$, the following operations can be attained by local computation at each party, and then the outcome can be assembled from the individual shares: 
\begin{enumerate}

\item[{\tt A1})] $\langle x \rangle + \langle y \rangle$ can be computed by $(x_1 + y_1, ..., x_n + y_n)$.

\item[{\tt A2})] $c \cdot \langle x \rangle$ can be computed by $(c \cdot x_1, ..., c \cdot x_n)$.

\item[{\tt A3})] $c + \langle x \rangle$ can be computed by $(c + x_1, x_2, ..., x_n)$.

\end{enumerate}

\subsubsection{Multiplication}
\

Given secretly shared $\langle x \rangle$ and $\langle y \rangle$, computing the product  $\langle x \rangle \cdot \langle y \rangle$ involves a given multiplication triple. A multiplication triple is defined by $(\langle a \rangle, \langle b \rangle, \langle c \rangle)$, where $a, b$ are some unknown random numbers and $c = a \cdot b$, are three secretly shared numbers already distributed among the parties. The triple is assumed to be prepared in a pre-processing phase. To compute  $\langle x  \rangle \cdot \langle  y \rangle$, it follows the below steps of operations ({\tt A4}):
\begin{enumerate}

\item[{\tt A4.1})] Compute $\langle\epsilon \rangle = \langle x  \rangle - \langle  a \rangle$ (by {\tt A1}). Then, reveal $\langle\epsilon \rangle$, which does not reveal $x$.

\item[{\tt A4.2})] Compute $\langle \delta \rangle = \langle y   \rangle - \langle  b \rangle$. Then, reveal $\langle\delta \rangle$.

\item[{\tt A4.3})] Finally, compute $\langle x  \rangle \cdot \langle  y \rangle = \langle c \rangle  + \epsilon  \cdot  \langle b \rangle +  \delta  \cdot  \langle a \rangle + \epsilon  \cdot  \delta$ (by {\tt A1}-{\tt A3}).

\end{enumerate}

\subsubsection{Message Authentication Code}
\

To safeguard against dishonest parties, who may perform incorrect computation, an information-theoretical message authentication code (MAC) can be used for verification. We write a MAC key as a global number $\widetilde{\alpha}$, which is unknown to the parties, and is secretly shared as $\langle \widetilde{\alpha} \rangle$. Every secretly shared number is encoded by a MAC as $\gamma(x) = \widetilde{\alpha} x$, which is secretly shared as $\langle \gamma(x) \rangle$. For each $\langle x \rangle$, each party $i$ holds a tuple $(x_i, \gamma(x)_i)$ and $\widetilde{\alpha}_i$, where $x = \sum_{i=1}^n x_i$, $\widetilde{\alpha} = \sum_{i=1}^n \widetilde{\alpha}_i$ and $\gamma(x) = \widetilde{\alpha} x = \sum_{i=1}^n  \gamma(x)_i$. If any party tries to modify her share $x_i$ uncoordinatedly, then he also needs to modify $\gamma(x)_i$ accordingly. Otherwise, $\gamma(x)$ will be inconsistent. However, it is difficult to modify $\gamma(x)_i$ without coordination among the parties, such that $\widetilde{\alpha} x = \sum_{i=1}^n  \gamma(x)_i$. Hence, it is possible to detect incorrect computation (possibly by dishonest parties) by checking the MAC.

To check the consistency of $x$, there is no need to reveal $\langle \widetilde{\alpha} \rangle$. One only needs to reveal $\langle x \rangle$, and then reveals $\widetilde{\alpha}_i - x \cdot \gamma(x)_i$ from each party $i$. One can check whether $\sum_{i=1}^n (\widetilde{\alpha}_i - x \cdot \gamma(x)_i) \overset{?}{=} 0$ for consistency. To prevent a dishonest party from modifying her share $x_i$ after learning other party's $x_j$. Each party needs to commit her share $x_i$ before revealing $x_i$ to others.

To maintain the consistency of MAC for operations {\tt A1}-{\tt A4}, the MAC needs to be updated accordingly as follows:
\begin{enumerate}

\item[{\tt B1})] $\langle x \rangle + \langle y \rangle$: Update MAC by $(\gamma(x)_1 + \gamma(y)_1, ..., \gamma(x)_n + \gamma(y)_n)$.

\item[{\tt B2})] $c \cdot \langle x \rangle$: Update MAC by $(c \cdot \gamma(x)_1, ..., c \cdot \gamma(x)_n)$.

\item[{\tt B3})] $c + \langle x \rangle$: Update MAC by $(c\cdot\alpha_1 + \gamma(x)_1, ..., c\cdot\alpha_n + \gamma(x)_n)$.

\item[{\tt B4})] $\langle x \rangle \cdot \langle y \rangle$: Update MAC at each individual step of {\tt A4.1}-{\tt A4.3} accordingly by {\tt B1}-{\tt B3}.

\end{enumerate}
The additions and multiplications of $\llangle  x \rrangle$ and $\llangle  y \rrangle$ follow {\tt A1}-{\tt A4} and the MACs will be updated accordingly by {\tt B1}-{\tt B4}. 

To verify the computation of a function, it only requires to check the MACs of the revealed values and the final outcome, which can be checked all efficiently together in a batch at the final stage by a technique of called ``random linear combination''.

\subsection{Pre-processing Phase}

In the pre-processing phase, all parties need to prepare a collection of triplets $(\langle  a \rangle, \langle b \rangle, \langle  c \rangle)$ where $c=a\cdot b$, each for a required multiplication operation.
Assume that the parties hold secretly shared numbers $a = \sum_{i=1}^N a_i$ and $b = \sum_{i=1}^N b_i$ (which has been generated by local random generation). Note that $a \cdot b= \sum_{i=1}^N a_i b_i + \sum_{i=1}^N \sum_{j=i"i\ne j}^N a_i b_j$. $a_i b_i$ can be computed locally. To distribute $a_i b_j$, one can use partial homomorphic cryptosystems, with encryption function ${\tt Enc}[\cdot]$ and decryption function ${\tt Dec}[\cdot]$ using party $i$'s public and private $(K_i^{\tt p}, K_i^{\tt p})$. First, party $i$ sends ${\tt Enc}_{K_i^{\tt p}}[a_i]$ to party $j$, who responds by $C_i = b_j {\tt Enc}_{K_i^{\tt p}}[a_i] - {\tt Enc}_{K_i^{\tt p}}[\tilde{c}_j]$, where $\tilde{c}_j$ is a random share generated by party $j$ and is encrypted by party $i$'s public key $K_i^{\tt s}$. Then party $i$ can obtain $\tilde{c}_j = {\tt Dec}_{K_i^{\tt s}}[C_j]$. Hence, $a_i b_j = \tilde{c}_i + \tilde{c}_j$, which are secret shares $a_i b_j$. The above generation assumes honest parties. To prevent cheating by dishonest parties, one would need to use proper zero-knowledge proofs before secret sharing \cite{cramer2015secure, dklpss13spdz}.

To generate a random mask $\llangle  r^i \rrangle$, each party $j$ needs to generate a random share $r^i_j$ locally. Then the parties follow the similar procedure of triplet generation to compute the secretly shared product $\langle  \gamma(r^i) \rangle$, where $\gamma(r^i) = \widetilde{\alpha} r^i$.

\subsection{Output and Validation Phase}
We describe random linear combination for batch checking. To check the MACs of a number of secretly shared numbers $\llangle  x^1 \rrangle, ..., \llangle  x^m \rrangle$ in a batch, first generate a set of random $({\tt r}^1, ..., {\tt r}^m)$. then reveal $\llangle  x^1 \rrangle, ..., \llangle  x^m \rrangle$. Each party $i$ computes $\sum_{j=1}^m {\tt r}^j (\widetilde{\alpha}_i - x^j \cdot \gamma(x^j)_i)$ and reveals it. All parties check whether $\sum_{i=1}^N \sum_{j=1}^m {\tt r}^j (\widetilde{\alpha}_i - x^j \cdot \gamma(x^j)_i) \overset{?}{=} 0$ for consistency in a batch checking.

\subsection{Protocol} \label{sec:spdz-protocol}

We summarize the SPDZ protocol as follows:
\begin{enumerate}

\item {\em Pre-processing Phase}: In this phase, a collection of shared random numbers will be constructed that can be used to mask the private input numbers. For each private input number of party $i$, there is a shared random number $\llangle r^i \rrangle$, where $r^i$ is revealed to party $i$ only, but not to other parties. All parties also prepare a collection of triplets $(\llangle  a \rrangle, \llangle b \rrangle, \llangle  c \rrangle)$ where $c=a\cdot b$, each for a required multiplication operation.

\item {\em Online Phase}: To secretly shares a private input number $x^i$ using $\llangle r^i \rrangle$, without revealing $x^i$, it proceeds as follows:
\begin{enumerate}

\item[1)] Party $i$ computes and reveals $z^i = x^i - r^i$ to all parties.

\item[2)] Every party sets $\llangle x^i \rrangle \leftarrow z^i + \llangle r^i \rrangle$.

\end{enumerate}
\smallskip

To compute an arithmetic circuit, implement the required additions or multiplications by {\tt A1}-{\tt A4} and the MACs are updated accordingly by {\tt B1}-{\tt B4}. 

\item {\em Output and Validation Phase}: All MACs will be checked for all revealed numbers and the final output value. It can check all in a batch using random linear combination. If there is any inconsistency in the MACs, then abort.

\end{enumerate}

Note that SPDZ cannot guarantee abort with fairness -- dishonest parties may learn some partial values, even when the protocol aborts. However, this is a fundamental problem for any multi-party computation protocol with a majority of dishonest users, where dishonest parties are not identifiable when the computation is aborted.
\section{Security Analysis} \label{sec:append4}

We adopt the most common approach of security analysis in cryptography, based on the {\em Ideal/Real-Model Simulation} paradigm to prove and formalize the security achieved by our protocols. We next briefly describe the simulation paradigm. The detailed explanation can be found in the tutorial \cite{Lindell2016HowTS}. 

In the {\em ideal} model, all the parties send their private inputs to a trusted third party, who performs the prescribed computations and outputs the results to each party. The security requirements are already satisfied in the ideal model.  The {\em real} model represents the realistic view of the privacy-preserving protocol. The security is defined by comparing what an adversary can learn in the real model to that in the ideal model. If what can be learned by an adversary in the real world can be totally simulated in the ideal world, then the adversary cannot learn more information in the real world than in the ideal world, we can say that a protocol $\Pi$ is as secure as its corresponding ideal functionality $\mathcal{F}$. We give a formal definition of the security of our protocol as below: 

\begin{customthm}{4}
Assuming the discrete logarithm problem underlying the Pedersen commitment scheme is hard and the non-interactive zero-knowledge proofs are secure with access to a random oracle, in the $\mathcal{F}_{\tt prep}$-hybrid model \cite{dklpss13spdz}, the protocol $\Pi_{\tt pess}$ securely implements ${\mathcal{F}_{\tt pess}}$ with abort in the presence of an adaptive, active adversary in a dishonest-majority setting, if for every probabilistic polynomial-time (PPT) adversary ${\mathcal{A}}$ in the real model, there also exists a PPT adversary ${\mathcal{S}}$ in the ideal model, such that for each $i \in N$:
$$\{{\tt IDEAL}_{{\mathcal{F}_{\tt pess}}, {\mathcal{S}}}^{{\mathcal{F}}_{\tt prep}}\} \overset{\rm comp}{\equiv} \{{\tt REAL}_{\Pi_{\tt pess}, {\mathcal{A}}}^{{\mathcal{F}}_{\tt prep}}\}\}$$
where ${\tt IDEAL}$ and ${\tt REAL}$ respectively refer to the views and outputs of the corrupted and honest users in both ideal and real worlds.

\end{customthm}

We sketch the proof of the above theorem. Our aim is to demonstrate a simulator in the ideal model that can create a computationally indistinguishable view from that of the adversary in the real model. Even with a different set of honest-users' inputs, the adversary should still be unable to tell computationally indistinguishable differences between the views. The simulator ${\mathcal{S}}$ {\em externally} interacts with the ideal functionality and {\em internally} runs a copy of the protocol $\Pi_{\tt pess} \diamond \mathcal{F}_{\tt prep}$ feeding messages to the adversary ${\mathcal{A}}$. However, it is a not trivial task for a simulator to emulate an adaptively malicious adversary, who is able to corrupt users at any time during the protocol. The challenge lies in the difficulty that it must produce a consistent view of the corrupted users throughout the protocol without knowing their inputs. Firstly, we define an ideal functionality ${\mathcal{F}}_{\tt a(t)}^{(1)}$ in the stage 1 {\em Pre-operation Scheduling} computing the total energy demands $a(t)$ before presenting the corresponding simulator ${\mathcal{S}}_{\tt a(t)}^{(1)}$.

\begin{longfbox}[border-break-style=none,border-color=\#bbbbbb,background-color=\#eeeeee,breakable=true,,width=\linewidth]
\begin{center}{Functionality ${\mathcal{F}}_{\tt a(t)}^{(1)}$}\end{center}

\noindent
{\bf Input:} On input ({\em input}, $U_i$, $a_i(t)$), the functionality stores $a_i(t)$.

\noindent
{\bf Output:} On input ({\em output}) from all honest users, the functionality computes and outputs $a(t)=\sum_{i=1}^N a_i(t)$ to all the users. 

\noindent
{\bf Abort:} On input ({\em abort}), the functionality outputs $\emptyset$.

\end{longfbox}

\bigskip

{\bf Initialize:} The simulator ${\mathcal{S}}$ first calls ${\mathcal{F}}_{\tt prep}$ to generate a sufficient number of multiplication triples and random numbers. Note that ${\mathcal{S}}$ has access to all the shares of the MAC key, random numbers and multiplication triples as it knows the decryption keys of public-key cryptosystem in the preprocessing phase. The adversary ${\mathcal{A}}$ firstly corrupts a set of users, denoted by ${\mathtt{C}}$. Then the adversary may adaptively make corruptions on other users during the protocol. Next, ${\mathcal{S}}$ produces $g, h=g^k \in {\mathbb Z}_p^*$, where $k=\log_g h$ is the trapdoor to Pedersen commitment, with which ${\mathcal{S}}$ is able to find out two pairs $(m, r), (m', r')$, such that ${\tt Cm}(m, r)={\tt Cm}(m', r')$. 

\smallskip

\begin{longfbox}[border-break-style=none,border-color=\#bbbbbb,background-color=\#eeeeee,breakable=true,,width=\linewidth]
\begin{center}{Simulator ${\mathcal{S}}_{\tt a(t)}^{(1)}$}\end{center}
\begin{enumerate}
\item For honest users $i \notin {\mathtt{C}}$, ${\mathcal{S}}$ will simply generates dummy inputs $\hat{a}_i(t) = 0, \hat{r}_i(t) \in Z_p$ and reveals a commitment $\hat{C}_i(t)={\tt Cm}\big(\hat{a}_i(t), \hat{r}_i(t)\big)$ with an ${\tt nzkpNN}[a_i(t)]$. For the corrupted users $i \in {\mathtt{C}}$, ${\mathcal{S}}$ can extract their inputs with the knowledge of all the shares of $\llangle r \rrangle$ and verifies the ${\tt nzkpNN}[a_i(t)]$.

{\em Remarks:} From the perspective of the adversary, the inputs of the honest users are indistinguishable from those in the real world due to the information-theoretically hiding properties of SPDZ secret-sharing (unless all the $N$ shares are collected, the inputs cannot be reconstructed), and of the Pedersen commitment. 

\item ${\mathcal{S}}$ firstly calls ${\mathcal{F}}_{\tt a(t)}^{(1)}$ to obtain the output $a(t)$. As ${\mathcal{S}}$ already computed an output $\hat{a}(t)$ using dummy inputs of the honest users, it can respectively modify the share and MAC of a random honest user by adding $a(t)-\hat{a}(t)$ and $\alpha \big(a(t)-\hat{a}(t)\big)$ with the MAC key $\alpha$ initialized in the preprocessing phase. Then ${\mathcal{S}}$ can perform the MAC check to evaluate and open $a(t)$. If the check passes, ${\mathcal{S}}$ calls ${\mathcal{F}}_{\tt a(t)}^{(1)}$ to output $a(t)$ to all the users. Otherwise, ${\mathcal{S}}$ sends {\tt Abort} to ${\mathcal{F}}_{\tt a(t)}^{(1)}$.

{\em Remarks:} No matter what inputs the adversary generates for the corrupted users, ${\mathcal{S}}$ can always create a computationally indistinguishable output distribution in the ideal model from that in the real model from the view of the adversary ${\mathcal{A}}$. For the evaluation of $a(t)$, each $i$-th share $\alpha_i a(t) - \gamma_i\big(a(t)\big)$ appears uniformly random to the adversary, which has exactly the same distribution in both ideal and real models. 

\end{enumerate}

\end{longfbox}

\smallskip

After the simulator provided the simulated input $\big(\hat{a}_i(t), \hat{r}_i(t), \hat{C}_i(t)\big)$ for $i \notin {\mathtt{C}}$, the adversary can corrupt an honest user $U_i$ at any time. As aforementioned, the simulator must reveal its entire internal states, including the inputs, shares of inputs and random values that are consistent with the commitment $\hat{C}_i(t)$ to simulate an adaptive adversary. It is easy to obtain the input $a_i(t)$ from ${\mathcal{F}}_{\tt a(t)}^{(1)}$. Regarding the random value, the simulator will take advantage of the trapdoor $k$ of the Pedersen commitment to obtain $r_i(t)=\hat{r}_i(t)+\big(\hat{a}_i(t)-a_i(t)\big) \cdot k^{-1}$, such that $\hat{C}_i(t)={\tt Cm}\big(a_i(t), r_i(t)\big)={\tt Cm}\big(\hat{a}_i(t), \hat{r}_i(t)\big)$. Moreover, $U_i$'s share of her initial dummy input $\hat{a}_i(t)$ is $\hat{a}_i(t)+r_i-r$. Thus, it is trivial for the simulator to reveal the share $a_i(t)+\hat{a}_i(t)-r+r_i$ by adding $a_i(t)$. (See online phase in Section \ref{sec:spdz-protocol}).

Next, we give a brief description of the SPDZ-based zero-knowledge proofs ${\tt zkpCm}[a_i(t)]$ in $\Pi_{\tt pess}^{(1)}$ and ${\tt nzkpSum}[{\tt Cost}_{\tt ess}, (P_i )_{i=1}^N]$ in $\Pi_{\tt pess}^{(2)}$. For ${\tt zkpCm}[a_i(t)]$, $z_{a_i(t)}$ and $z_{r_i(t)}$ are collectively computed by all users and will be evaluated via MAC check to prove their correctness. A similar simulator to ${\mathcal{S}}_{\tt a(t)}^{(1)}$ can be constructed to emulate the ideal functionality computing $z_{a_i(t)}$ and $z_{r_i(t)}$. The challenge $\beta(t)$ is uniformly random independent of the prover's input as it is obtained by summing the random values generated by all the users. Thus, this zero-knowledge proof is secure given the proof of completeness, soundness, zero-knowledge properties in Section \ref{sec:zkpproofs}. The same security argument applies to ${\tt nzkpSum}[{\tt Cost}_{\tt ess}, (P_i )_{i=1}^N]$.

We skip the details for the simulator ${\mathcal{S}}_{\tt Cost_{\tt ess}}^{(2)}$ emulating ideal functionality computing the total payment ${\tt Cost}_{\tt ess}=\sum_i^N P_i$ as it is similar to ${\mathcal{S}}_{\tt a(t)}^{(1)}$ except using a different input $P_i$.

\section{Ethereum Blockchain Platform \& Smart Contracts} \label{sec:smartcontract}

In this section, we provide a brief description of {\em Ethereum blockchain platform} and {\em Solidity programming language} as well as the details on the implementations of the smart contracts in our protocols. 

\subsection{Background}

Bitcoin was the first widely adopted digital currency on a permissionless distributed ledger. Bitcoin relies on a tampering-resistant ledger based on cryptographic signatures. Tampering-resistance ensures integrality when the ledger is maintained by a network of peer-to-peer systems called ``miners''. The miners are incentivized by cryptocurrency rewards for updating and validating the transaction records. Since the distributed ledger can be modified by multiple systems simultaneously, it is crucial to ensure consistency by a distributed consensus protocol among untrusted peer-to-peer systems, based on proof-of-work (by solving computational puzzles) or proof-of-stake (by demonstrating ownership of digital assets). 

Subsequently, Ethereum was built on the Bitcoin ideas by expanding its functions to support general computing as smart contracts along with transactions. Bitcoin operates using a transaction-output-based system, called unspent transaction outputs (UTXOs), whereas Ethereum operates using accounts and balances in a manner called state transitions. Smart contracts, which are code programmed in high-level logic, will be compiled into byte code and executed in the virtual machine of miners. Miners will charge additional crytocurrency payments called gas costs, because the extra computational tasks incurred by smart contracts will be broadcast throughout the blockchain. Smart contracts are implemented in a high-level programming language, such as Solidity \cite{solidity}.
 
It is worth noting that Bitcoin and Ethereum were only supposed to enable decentralization, but do not ensure privacy. In fact, the transaction histories of many crytocurrencies are visible to the public. There are certain high-profiled prosecution of darknet operators based on the evidence of Bitcoin transactions. Supporting privacy in blockchain is a crucial on-going research topic.

\subsection{Smart Contract Implementation}

We next explain how {\em Multi-Signature} smart contract can achieve the step (5) and (6) of the stage {\em Cost-sharing Payment} by the following methods:

\begin{enumerate}
\item{\tt submitTransaction().} This method allows each user to submit the ${\tt zkpSum}[{\tt Cost}_{\tt ess}, (P_i )_{i=1}^N]$ that they have agreed upon off the chain. The method will compare whether users have submitted the same ${\tt zkpSum}$. 

\item{\tt confirmTransaction().} On one hand, this method allows each user to confirm that the stored ${\tt zkpSum}$ in the smart contract is the one that they have agreed upon off the chain. On the other, each user is required to submit a ${\tt nzkpNN}[{\tt Bal}({\tt ad}_i) - P_i]_{i=1}^N$, which will be validated to prove that there is sufficient balance in his account to pay for the energy cost.

\item{\tt executeTransaction().} This method can only be executed by the operator unless all the users have already confirmed the transaction. The method will validate the ${\tt zkpSum}$ before calling {\em ESToken} smart contract to credit ${\tt Cost}_{\tt ess}$ to the operator's account and debit the corresponding payment from each user's account. 

\item{\tt secretlyJointTransfer().} This method, defined within the {\em EStoken} smart contract is invoked by {\tt executeTransaction()}, which actually performs the real transfer between multiple accounts. 
\end{enumerate}

\end{document}